\documentclass[3p, preprint, sort&compress]{elsarticle}
\usepackage{soul, xcolor}

\usepackage{caption,subcaption}
\usepackage{amsfonts,bbold}
\usepackage{graphicx}
\usepackage{amssymb,amsmath}
\usepackage{hyperref} 		% % [hidelinks]

\usepackage{mathtools}

\renewcommand{\d}{\text{d}}
\renewcommand{\H}{\mathcal{H}}
\renewcommand{\O}{\mathcal{O}}

\newcommand{\M}{\text{M}}
\newcommand{\m}{\text{m}}
\newcommand{\T}{\text{T}}

\newcommand{\gate}{\text{gate}}
\newcommand{\lab}{\text{lab}}
\newcommand{\rot}{\text{rot}}
\renewcommand{\max}{\text{max}}
\newcommand{\RWA}{\text{RWA}}
\newcommand{\eff}{\text{eff}}

\makeatletter
\newcommand{\specialcell}[1]{\ifmeasuring@#1\else\omit$\displaystyle#1$\ignorespaces\fi}
\makeatother

\makeatletter
\newcommand*\bigcdot{\mathpalette\bigcdot@{.5}}
\newcommand*\bigcdot@[2]{\mathbin{\vcenter{\hbox{\scalebox{#2}{$\m@th#1\bullet$}}}}}
\makeatother

\newcommand{\dersub}[2]{\overset{\,\,\scriptscriptstyle{(\hspace{-.8pt}\raisebox{0.8pt}{\text{\large{.}}} \hspace{-.8pt})^{#2}}}{#1}}
\newcommand{\dersubLong}[2]{\hspace{-8pt}\overset{\,\,\,\,\,\,\,\,\scriptscriptstyle{(\hspace{-0.8pt}\raisebox{0.8pt}{\text{\large{.}}} \hspace{-0.8pt})^{#2}}}{#1}}

\DeclareRobustCommand{\cev}[1]{\text{\reflectbox{$\vec{#1}$}}}

\DeclareRobustCommand{\prodIn}{%
  \mathord{\vphantom{\rightarrow}\text{%
    \ooalign{%
      \,\,\,\,\raisebox{-1em}{\scalebox{3.2}[2]{$\cev{} $}}\kern+0.05em\raisebox{0.07ex}{\scalebox{0.88}{$ \ $}}\cr
      \kern0.02em\raisebox{0.0em}{$\prod$}\cr
    }%
  }}%
}

\DeclareRobustCommand{\prodEqn}{%
  \mathord{\vphantom{\rightarrow}\text{%
    \ooalign{%
      \,\,\,\,\,\raisebox{-1.25em}{\scalebox{4.3}[2.5]{$\cev{} $}}\kern+0.05em\raisebox{0.07ex}{\scalebox{0.88}{$ \ $}}\cr
      \kern0.02em\raisebox{0.0em}{$\prod_{j=0}^{n}$}\cr
    }%
  }}%
}

\DeclareMathOperator*{\op}{%
\mathchoice%
{\ooalign{\raisebox{.05\height}{$\displaystyle\prod$}\cr\hidewidth{$\displaystyle\longleftarrow$}\cr\hidewidth{$\displaystyle\longleftarrow$}\cr\hidewidth{$\displaystyle\phantom{\prod}$}\hidewidth\cr}}
  {\ooalign{\scalebox{.9}{$\textstyle\prod$}\cr\hidewidth\scalebox{.8}{$\!\textstyle\longleftarrow$}\hidewidth\cr}}
  {\ooalign{\raisebox{.1\height}{\scalebox{.75}{$\scriptstyle\prod$}}\cr\hidewidth$\scriptstyle\leftarrow$\cr}}
  {\ooalign{\raisebox{.1\height}{\scalebox{.75}{$\scriptstyle\prod$}}\cr\hidewidth$\scriptstyle\leftarrow$\cr}}
}

\begin{document}

\author[1]{Daniel Zeuch\corref{cor1}}
\ead{dzench@gmail.de (sic)}

\author[2]{Fabian Hassler}
\author[3]{Jesse J. Slim}
\author[1,2]{David P. DiVincenzo}

\cortext[cor1]{Corresponding author}
\address[1]{Peter Gr\" unberg Institut:  Theoretical Nanoelectronics, Research Center J\" ulich, 52428 J\"ulich, Germany}
\address[2]{Institute for Quantum Information, RWTH Aachen University, 52062 Aachen, Germany}
\address[3]{Center for Nanophotonics, AMOLF, Science Park 104, 1098 XG Amsterdam, The Netherlands}

\date{\today}

\title{Exact Rotating Wave Approximation}

\begin{abstract}
The Hamiltonian of a linearly driven two-level system, or qubit, in the standard rotating frame contains non-commuting terms that oscillate at twice the drive frequency, $\omega$, rendering the task of analytically finding the qubit's time evolution nontrivial.  The application of the rotating wave approximation (RWA), which is suitable only for drives whose amplitude, or envelope, $H_1(t)$, is small compared to $\omega$ and varies slowly on the time scale of $1/\omega$, yields a simple Hamiltonian that can be integrated relatively easily.  We present a series of corrections to the RWA Hamiltonian in $1/\omega$, resulting in an effective Hamiltonian whose time evolution is accurate also for time-dependent drive envelopes in the regime of strong driving, i.e., for $|H_1(t)| \lesssim \omega$.  By extending the Magnus expansion with the use of a Taylor series we introduce a method that we call the Magnus-Taylor expansion, which we use to derive a recurrence relation for computing the effective Hamiltonian.  We then employ the same method to derive kick operators, which complete our theory for non-smooth drives.  The time evolution generated by our kick operators and effective Hamiltonian, both of which depend explicitly on the envelope and its time derivatives, agrees with the exact time evolution at periodic points in time.  For the leading Hamiltonian correction we obtain a term proportional to the first derivative of the envelope, which competes with the Bloch-Siegert shift.  
\end{abstract}


\maketitle
\tableofcontents

\section{Introduction}
\label{introduction}

Coherent driving of quantum systems plays a central role in many areas of physics and chemistry \cite{grifoni98}.  The specific task with arguably the highest demands on the accuracy of the desired operations carried out by such driving is the manipulation of two-level systems, or qubits, that form the basic elements of a quantum computer \cite{nielsen10, zagoskin11}.  

Perhaps the simplest abstraction of the interaction between light and matter is formalized by the semiclassical Rabi model \cite{rabi37}, which can be used to describe a qubit  interacting with a classical, circularly polarized drive.  We study the related and, from a theoretical point of view, significantly richer problem of a qubit subject to a \emph{linearly} polarized drive, which has been considered early on by Bloch and Siegert \cite{bloch40}.  Denoting the qubit and drive frequencies by $\omega_0$ and $\omega$, respectively, the system Hamiltonian in the laboratory frame reads ($\hbar = 1$)
\begin{equation}
	\mathcal H_{\lab}(t) = \frac{\omega_0}2 \sigma_z + \frac{H_1(t)}{2} \cos(\omega t + \phi) \sigma_x,
	\label{Hlab}
\end{equation}
where $H_1(t)$ is the time-dependent amplitude function, or envelope, of the applied drive.  We denote the Pauli matrices by $\sigma_i$ with $i = x$, $y$ and $z$, and assume that the drive has a constant phase offset, $\phi$.  In our study we work in a suitable frame of reference, which rotates around the $z$ axis with the frequency $\omega$ of the applied field.  

The minimal time required to effectively manipulate the state of a qubit by such driving is inversely proportional to the drive strength $H_1(t)$.  One thus needs a strong drive for fast pulses, which are often desired because they allow a large number of operations to be carried out within the coherence time of the qubit.  The ratio of the amplitude $H_1(t)$ and the qubit resonance frequency $\omega$ can be used to distinguish between different parameter regimes.  The regime perhaps best understood is that for which the drive amplitude is both constant in time and small compared to the drive frequency, $H_1/\omega \ll 1$, also known as the weak coupling regime.  In this case, resonant driving ($\omega = \omega_0$) results in Rabi oscillations with frequency proportional to $H_1$.  In the present study, we are concerned with the strong coupling regime in which $H_1/\omega \lesssim 1$, and place special interest in the consequences on the qubit's time evolution due to a time-varying field strength $H_1 = H_1(t)$.  

The use of shaped pulses is common for the manipulation of quantum systems through coherent driving.  It has a long history in nuclear magnetic resonance \cite{mcdonald91, freeman91}, in which implementations of non-square envelope functions have already been reported in the late 1970s \cite{sutherland78}.  In quantum information processing, searching for suitable pulse shapes for quantum gates (including single-qubit rotations) falls under the topic of optimal control theory \cite{glaser15}.  Such a search may be streamlined by the ability to predict the driven qubit's time evolution using a high-precision approximation that is easy to integrate numerically.  Our work provides such an approximation for weak to strong driving as defined above.  

In the standard rotating frame the Hamiltonian of a linearly-driven qubit contains non-commuting terms that oscillate quickly, at twice the drive frequency \cite{slichter63}.  If these terms, which can be attributed to the counter-rotating field of the drive, are fully taken into account, there is no simple analytic form for the qubit's exact time evolution.  If the drive is weak ($|H_1(t)|\ll\omega$), near resonant ($\omega\approx \omega_0$), and varies only slightly on the time scale of the inverse qubit frequency $1/\omega$, the application of the rotating wave approximation (RWA) yields a simple Hamiltonian which is straightforward to integrate \cite{cohen98}.  However, for moderately strong field strengths $|H_1(t)|/\omega \gtrapprox 0.01$ the RWA is no longer applicable for many quantum-information related applications, and corrections that scale as some power in $1/\omega$, such as the well-known Bloch-Siegert shift \cite{bloch40}, may be used to improve the accuracy of the predicted time evolution.  

The problem of a periodically driven two-level system has been studied using the Magnus expansion,\footnote{We note that Ref.~\cite{rau98} investigates the problem of a spin in a time-dependent magnetic field using an approach related to the Magnus expansion.  In that work the time evolution operator is written as a product of three consecutive rotations about mutually perpendicular axes resulting in differential equations that are solved perturbatively.} which provides a means of performing time-dependent perturbation theory at the Hamiltonian level in which unitarity of the time evolution is inherently preserved \cite{magnus1954, ernst87, waugh07}.  The dressed-state formalism \cite{cohen73} has also been employed in a study concerning the time evolution of a periodically driven multi-level system \cite{yang17}.  Most work on the time evolution under periodic driving of quantum systems, however, is based on Floquet theory, starting with an influential analysis by Shirley in 1965 \cite{shirley65}.  While the basic formulation of Floquet's theorem may be used directly in such an investigation \cite{aravind84, mananga11, schmidt18}, the more common Floquet approach is to obtain an infinite dimensional, time-independent Hamiltonian by using an extended Hilbert space \cite{shirley65, sambe73}, and then introduce a two-time formalism that allows the formal separation of a micromotion and an effective, coarse-grained evolution  \cite{sambe73, howland74, breuer89ZPD, peskin93, drese99, novicenko17}.  

The first time that the Magnus expansion, one of the main methods used in the present investigation, has been applied to coherent driving of quantum systems has been in the context of nuclear magnetic resonance \cite{haeberlen68, evans68} (see also Ref.~\cite{feldman84}).  This paved the path for what is known as average Hamiltonian theory \cite{haeberlen68, waugh68}, in which \emph{effective Hamiltonians} are used to approximate the time evolution of the driven system.  In the meantime, the Magnus expansion has also been combined with the Floquet approach \cite{casas01, blanes09, mananga11, bukov2015universal}; this introduced novel concepts such as \emph{kick operators} \cite{mananga11, bukov2015universal} (see also Ref.~\cite{goldman14}) and a \emph{gauge degree of freedom} of effective Hamiltonians \cite{bukov2015universal}.  A recent study determines the \emph{stroboscopic time evolution} via a Magnus expansion for a driven qubit using a frequency chirp \cite{nalbach18}.  Many of these concepts play an important role in the present work.  

Most previous work that aims at determining the time evolution operator for the driven qubit has assumed periodic Hamiltonians \cite{shirley65, sambe73, aravind84, feldman84, breuer89ZPD, casas01, blanes09, mananga11, yang17, schmidt18, goldman14, bukov2015universal}.  While several analyses \cite{yang17, peskin93, drese99, goldman14, bukov2015universal, nalbach18} explore some consequences due to adiabatic changes of drive parameters, Ref.~\cite{novicenko17} explicitly determines a time evolution operator assuming a nonzero first derivative of these drive parameters.  However, realistic drives are often turned on non-adiabatically, and there may be substantial effects due to a nontrivial time dependence of the drive envelope.  Such complications are of particular importance for strong or shaped pulses, e.g., to minimize leakage out of the computational space via DRAG shaping for superconducting qubits \cite{motzoi09}, or to increase the fidelity of gates for singlet-triplet qubits \cite{cerfontaine14, varvelis19}.  An investigation that allows for relatively generic pulse amplitudes is that of Ref.~\cite{wu07}, where a recursive procedure for obtaining the wave function of the driven qubit is developed; this investigation, however, has been conducted for ultra-strong driving with $|H_1(t)| \gg \omega$, and thus for parameters that lie outside the range considered here.  Reference \cite{barnes12} establishes a formalism for reverse-engineering drive functions that result in certain qubit trajectories.  This survey makes it evident that the problem of strictly periodic driving has been discussed extensively, while relatively little attention has been devoted to the problem considered here, i.e., determining the time evolution for the system governed by the Hamiltonian (\ref{Hlab}) for a wide class of time-dependent amplitude functions in the strong coupling regime.  

We note that in Ref.~\cite{giscard15}, Giscard \textit{et al.}~introduced the path-sum method, which solves the Volterra integral equation of the second kind using a Neumann series in a way that the solution can be obtained at low cost, requiring only a finite number of computation steps.  Schroedinger's equation is a special case of the Volterra equation, and so this path-sum method is applicable to a wide variety of systems.  An application of this method to our driven qubit system (for a constant drive amplitude) has recently been demonstrated in Ref.~\cite{giscard2019general}.  A noteworthy difference between the path-sum method and our perturbative effective Hamiltonian method is that only in our case the time evolution operator is always unitary when breaking the perturbation series at finite order---this is a direct consequence of our usage of the Magnus expansion.  

To approach the driven qubit problem described above, we introduce the \emph{Magnus-Taylor expansion}, a new method for time-dependent perturbation theory [see Sec.~\ref{MT}].  When considering, for example, the rotating-frame Hamiltonian that corresponds to Eq.~(\ref{Hlab}), this method combines a Magnus expansion with a Taylor series of the amplitude function $H_1(t)$ in a way that allows us to evaluate the integrals occurring in the Magnus expansion asymptotically.  Using this Magnus-Taylor expansion, we derive our main results: a time-dependent effective Hamiltonian and associated kick operators, each given as a series expansion in the inverse drive frequency.  This Hamiltonian, in combination with the kick operators, generates what we call an effective time evolution that agrees with the exact time evolution at periodic, or stroboscopically-defined, points in time.  

Our effective Hamiltonian, denoted $\H_{\eff}$, explicitly depends not only on the envelope $H_1(t)$, but also on its time derivatives $\dot H_1(t)$, $\ddot H_1(t)$, \ldots, all of which are assumed to change only slightly over the period of the drive.  To be precise, in our theory $\H_\eff (t)$ is an explicit {\em local} function of the envelope and its derivatives, i.e., 
\begin{equation}
	\H_\eff(t)=f(H_1(t),\dot H_1(t),\ddot H_1(t), \ldots).
	\label{HEffDD}
\end{equation}
We took inspiration in this from the Local Density Approximation of the Density Functional Theory for electronic structure.  Our central assumption on a slowly varying, weak drive amplitude can be formulated as follows:  for all times $t$ we require
\begin{equation}
	|\dersub{H_1}{k}(t)| \lesssim \omega^{k+1} \qquad \forall k \in \mathbb{N}_0,
	\label{assumption}
\end{equation}
where $\dersub{H_1}{k}$ denotes the $k$th derivative of the envelope.  While this assumption seems to considerably limit the applicability of our theory, below in Sec.~\ref{intro:gauge_kicks} we introduce the role of kick operators that are used to account for more realistic drives.  From Eq.~(\ref{HEffDD}) it is clear that $\H_{\eff}$ has only a slow time dependence compared to the rotating-frame Hamiltonian, which, as noted above, contains terms oscillating at frequency $2\omega$.  The effective Hamiltonian is therefore, similar to the RWA Hamiltonian, relatively easy to integrate numerically.  At the same time, by computing the effective Hamiltonian series up to appropriate order in $1/\omega$, our method allows one to determine the stroboscopic time evolution up to any desired accuracy.  Since our approach combines the advantageous features of the RWA with the ability to achieve arbitrarily high accuracy, we call our effective Hamiltonian method the ``exact'' rotating wave approximation.  

The remainder of this paper is organized as follows.  In what remains of the Introduction, we describe our anticipated solution to the driven qubit problem.  Section \ref{RWA} sets the stage by transforming the Hamiltonian to the standard rotating frame.  Most importantly, Secs.~\ref{HamiltoniansEff} and \ref{stroboscopicTimeEvolution} clarify how our concept of effective Hamiltonians is fundamentally tied to the idea of a stroboscopic time evolution.  In Sec.~\ref{intro:gauge_kicks} we introduce a gauge degree of freedom, an inherent aspect our effective Hamiltonian theory, together with kick operators, which extend our theory for many realistic drive envelopes.  In Sec.~\ref{effectiveH} we derive the recursive procedure that yields the desired Hamiltonian (\ref{HEffDD}) as a series expansion in $1/\omega$.  To do this, we first apply the Magnus expansion to the driven-qubit problem [Sec.~\ref{M}], next introduce the Magnus-Taylor expansion as a new tool for time-dependent perturbation theory [Sec.~\ref{MT}], and then apply this tool to the problem at hand to derive the central recurrence relation generating our effective Hamiltonian [Sec.~\ref{procedure}].  We also present an exemplary calculation of an effective Hamiltonian [Sec.~\ref{HeffExample}], and derive a simplified computation method for effective Hamiltonians assuming a constant drive envelope [Sec.~\ref{Heffconst}].  Deploying the Magnus-Taylor expansion a second time, in Sec.~\ref{nonanalytic} we derive the kick operators as a series expansion in $1/\omega$, and conclude in Sec.~\ref{conclusions}.  \ref{appendix:examples} contains explicit results for our effective Hamiltonians up to second order in $1/\omega$.  

\subsection{Rotating Frame}
\label{RWA}

We shift our discussion from the laboratory frame to the standard rotating frame, which is associated with the drive frequency $\omega$---our entire discussion below assumes this rotating frame of reference.  The corresponding Hamiltonian, defined by the usual transformation $\mathcal H_{\rot} = \tilde U^{\dagger} \mathcal H_{\lab} \tilde U - i \tilde U^\dagger \frac{\partial}{\partial t}\tilde U$ with $\H_{\lab}$ given in Eq.~(\ref{Hlab}) and $\tilde U(t) = e^{-i \omega t \sigma_z/2}$ \cite{messiah1964quantum}, evaluates to
\begin{equation}
	\mathcal H_{\rot}(t) = \frac{H_1(t)}4 ( \cos(\phi)\sigma_x + \cos(2\omega t + \phi) \sigma_x + \sin(\phi)\sigma_y - \sin(2\omega t + \phi)\sigma_y) + \frac\Delta2 \sigma_z.
	\label{Hrot0}
\end{equation}
Here we have introduced the detuning $\Delta = \omega_0 - \omega$.  Note that after this transformation the period of the counter-rotating field of the drive (henceforth:  the drive) is
\begin{equation}
	 \qquad \qquad \qquad \qquad \qquad t_c = \pi/\omega, \qquad \qquad \qquad \text{(rotating frame).}
	\label{tc}
\end{equation}
For simplicity, when considering examples we often restrict our discussion to drives that are in resonance with the qubit, i.e., $\omega = \omega_0$, or $\Delta = 0$, and that have zero phase offset, $\phi = 0$.  For this special case the rotating-frame Hamiltonian reduces to
\begin{equation}
	\mathcal H_{\rot}(t) = \frac{H_1(t)}4 ( \sigma_x + \cos(2\omega t) \sigma_x - \sin(2\omega t)\sigma_y), \quad \qquad (\Delta=0, \ \phi=0).
	\label{Hrot2}
\end{equation}

Our goal is to determine the time evolution of the driven qubit.  The problem at hand is thus to solve the time-dependent Schroedinger equation in the rotating frame,
\begin{equation}
	i (\partial/\partial t) |\psi(t)\rangle = \mathcal H_{\rot}(t) |\psi(t)\rangle,
	\label{schroedinger}
\end{equation}
whose formal solution is $|\psi(t_f)\rangle = U(t_f, t_i) |\psi(t_i)\rangle$ for some initial and final times $t_i$ and $t_f$, respectively.  Here the time evolution operator takes the usual form 
\begin{equation}
	U(t_f, t_i) = \mathcal{T}e^{-i\int_{t_i}^{t_f} \d \tau \mathcal H_{\rot}(\tau)},
	\label{Ugeneric}
\end{equation}
where $\mathcal{T}$ is the time ordering operator.  Note that here, given that the Hamiltonian (\ref{Hrot0}) does not commute with itself at different times, the evaluation of this time-ordered product is nontrivial even in the simplest case of a constant drive envelope.

\subsection{Effective Hamiltonians}
\label{HamiltoniansEff}

In the weak coupling limit, defined by a small drive amplitude $|H_1(t)|\ll \omega$, it is justified to apply the rotating wave approximation \cite{scully99} (or RWA).  To do this, one neglects the fast-oscillating terms in the rotating-frame Hamiltonian, resulting in a significantly simpler Hamiltonian that depends on time solely through the amplitude function $H_1(t)$,
\begin{eqnarray}
	\qquad \qquad \qquad
	\mathcal H_{\RWA}(t) & = & \frac{H_1(t)}4 ( \cos(\phi)\sigma_x + \sin(\phi)\sigma_y) + \frac\Delta2 \sigma_z
	\label{HRWAgeneric} \\
			& = & \frac{H_1(t)}4 \sigma_x, \qquad\qquad\qquad\qquad \qquad \qquad \qquad (\Delta = 0, \ \phi = 0).
	\label{HRWA}
\end{eqnarray}
If $\Delta = 0$ and, for example, the qubit state is initialized to $|\psi(t=0)\rangle=|0\rangle$, the RWA trajectory of $|\psi(t)\rangle$ obtained by solving Eq.~(\ref{schroedinger}) results in Rabi oscillations (with period $4\pi/H_1$ for constant $H_1$).  As noted above, for field strengths $|H_1(t)| \gtrsim 0.01\omega$ the RWA if often not applicable.  The effective Hamiltonian  introduced in this paper generalizes the RWA Hamiltonian in the sense that it can be used to approximate the exact trajectory for strong drive strengths up to $|H_1(t)| \lesssim \omega$.  

A central feature of our effective Hamiltonian is that it generates a stroboscopic time evolution formalized and exemplified below in Sec.~\ref{stroboscopicTimeEvolution}.  We describe this evolution as being ``stroboscopic'' because it agrees with the time evolution of the exact Hamiltonian at points equally spaced in time,
\begin{equation}
	\{\ldots t_0 - 2t_c, t_0 - t_c, t_0, t_0 + t_c, t_0 + 2t_c, \ldots \}.
	\label{strobo}
\end{equation}
Here the spacing is equal to the drive period in the rotating frame, $t_c=\pi/\omega$ [cf.~Eq.~(\ref{tc})], and the constant time offset is chosen to be $t_0\in[0, t_c)$.  The two cases of time-independent and time-dependent drive envelopes are qualitatively different.  

\subsubsection{Time-Independent Drive Envelope}
\label{intro:Time Independent Drives}

Let us first consider the case of constant $H_1(t) = H_1$.  As noted above, usage of the RWA is not always justified if the amplitude $H_1$ is an appreciable fraction of the drive frequency $\omega$.  A better, systematic approximation for the time evolution can be obtained by our effective Hamiltonian for constant drive envelopes, which depends only on the time offset $t_0$ defining the set of stroboscopic points (\ref{strobo}).  We introduce this Hamiltonian as a series expansion in $1/\omega$,
\begin{equation}
	\H_{\eff}(t_0) = \sum_{k=0}^{\infty} \frac{h_{k}(t_0)}{\omega^k},
	\label{Heff0}
\end{equation}
which, being independent of the current time $t$, allows for a simple evaluation of the time evolution operator (\ref{Ugeneric}).  Note that this lack of dependence on time $t$ follows directly from $H_1(t) = H_1$ combined with the fact that the effective Hamiltonian's time dependence is solely through the amplitude function [cf.~Eq.~(\ref{HEffDD})].  

We exemplify some qualitative features beyond the RWA for the simple case of resonant driving ($\Delta =0$) with zero phase offset ($\phi=0$), in which the system is governed by the Hamiltonian (\ref{Hrot2}).  The effective Hamiltonian, which can be computed using the method derived in Sec.~\ref{effectiveH}, given up to seventh order in $1/\omega$ and setting, for simplicity, $t_0 = 0$, reads
\begin{eqnarray}
	\H_{\eff}(t_0=0)&=&\frac{H_1}4 \sigma_x - \frac{H_1^2}{32\omega} \sigma_z -\frac{H_1^3}{256 \omega^2} \sigma_x -\frac{H_1^4 }{512 \omega ^3}\sigma_z-\frac{3 H_1^5}{8192 \omega ^4} \sigma_x - \frac{61 H_1^6 }{786432 \omega ^5}\sigma_z \nonumber \\ 
		&& -\frac{341 H_1^7 }{12582912 \omega ^6} \sigma_x -\frac{937 H_1^8}{1811939328 \omega ^7}\sigma_z + \mathcal O(1/\omega^8).
	\label{Hww-2}
\end{eqnarray}
Here, the first term is the RWA Hamiltonian given in Eq.~(\ref{HRWA}), while all other terms are corrections beyond the RWA.  The first-order contribution, $-(H_1^2/32\omega) \sigma_z$, is the well known Bloch Siegert shift \cite{bloch40}, which indicates a shift in the qubit resonance frequency on account of its proportionality to $\sigma_z$.  Conversely, the correction terms proportional to $\sigma_x$ indicate a decrease in the effective driving strength, or the Rabi frequency.

\subsubsection{Time-Dependent Drive Envelope}

The main purpose of this paper is to develop a theory for computing effective Hamiltonians similar to that given in Eq.~(\ref{Hww-2}), but for time-dependent envelopes $H_1(t)$.  Consistent with a $1/\omega$ expansion we assume the drive satisfies Eq.~(\ref{assumption}), which states that the absolute values of the envelope and its derivatives are small with respect to $\omega$.  This effective Hamiltonian, in contrast to the Hamiltonian (\ref{Heff0}), depends not only on the time offset $t_0$ but also on the current time $t$,
\begin{equation}
	\H_{\eff}(t; t_0) = \sum_{k=0}^{\infty} \frac{h_{k}(t; t_0)}{\omega^k}.
	\label{Heff1}
\end{equation}

An example effective Hamiltonian up to order $1/\omega^2$ for the case of resonant driving and zero phase offset, or $\Delta=0$ and $\phi=0$, together with $t_0=0$ is given by
\begin{equation}
	\H_{\eff}(t; t_0=0) = \frac{H_1(t)}{4} \sigma_x - \frac{H_1(t)^2}{32\omega} \sigma_z + \frac{\dot H_1(t)}{8\omega} \sigma_y - \frac{H_1(t)^3}{256 \omega^2} \sigma_x + \frac{\ddot H_1(t)}{16\omega^2} \sigma_x + \mathcal O(1/\omega^3),
	\label{Hww-3}
\end{equation}
which, in accordance with Eq.~(\ref{HEffDD}), is an explicit function of the derivatives $\dot H_1$ and $\ddot H_1$ as well as $H_1$.  When comparing the Hamiltonian (\ref{Hww-3}) with that given in Eq.~(\ref{Hww-2}), we find that for the lowest-order correction $\sim 1/\omega$ there is, besides the Bloch-Siegert shift, a term proportional to $\dot H_1(t)$.  We note that Hamiltonian terms proportional to the first and second derivatives of the envelope have already been found in Refs.~\cite{motzoi12, motzoi13}; however, these terms differ from those presented here by both their derivation and purpose.  

Generic examples of effective Hamiltonians for variable $t_0$, $\Delta$ and $\phi$ are presented in \ref{appendix:examples}.  Consulting, for example, the effective Hamiltonian (\ref{leadingCorrection}) for off-resonant driving with $\Delta \neq 0$, we find yet another first-order correction, which is proportional to $\Delta\,H_1$.  Finally, as expected, for constant $H_1(t) = H_1$ the effective Hamiltonian (\ref{Hww-3}) reduces to that for time-independent envelopes given above in Eq.~(\ref{Hww-2}).

%Consulting \ref{appendix:examples}, one can find yet another first-order correction proportional to $\Delta\,H_1$ for the case of more generic, off-resonant driving [see, e.g., Eq.~(\ref{leadingCorrection})].  

\subsection{Stroboscopic Time Evolution}
\label{stroboscopicTimeEvolution}

\begin{figure}
	\includegraphics[width = \columnwidth]{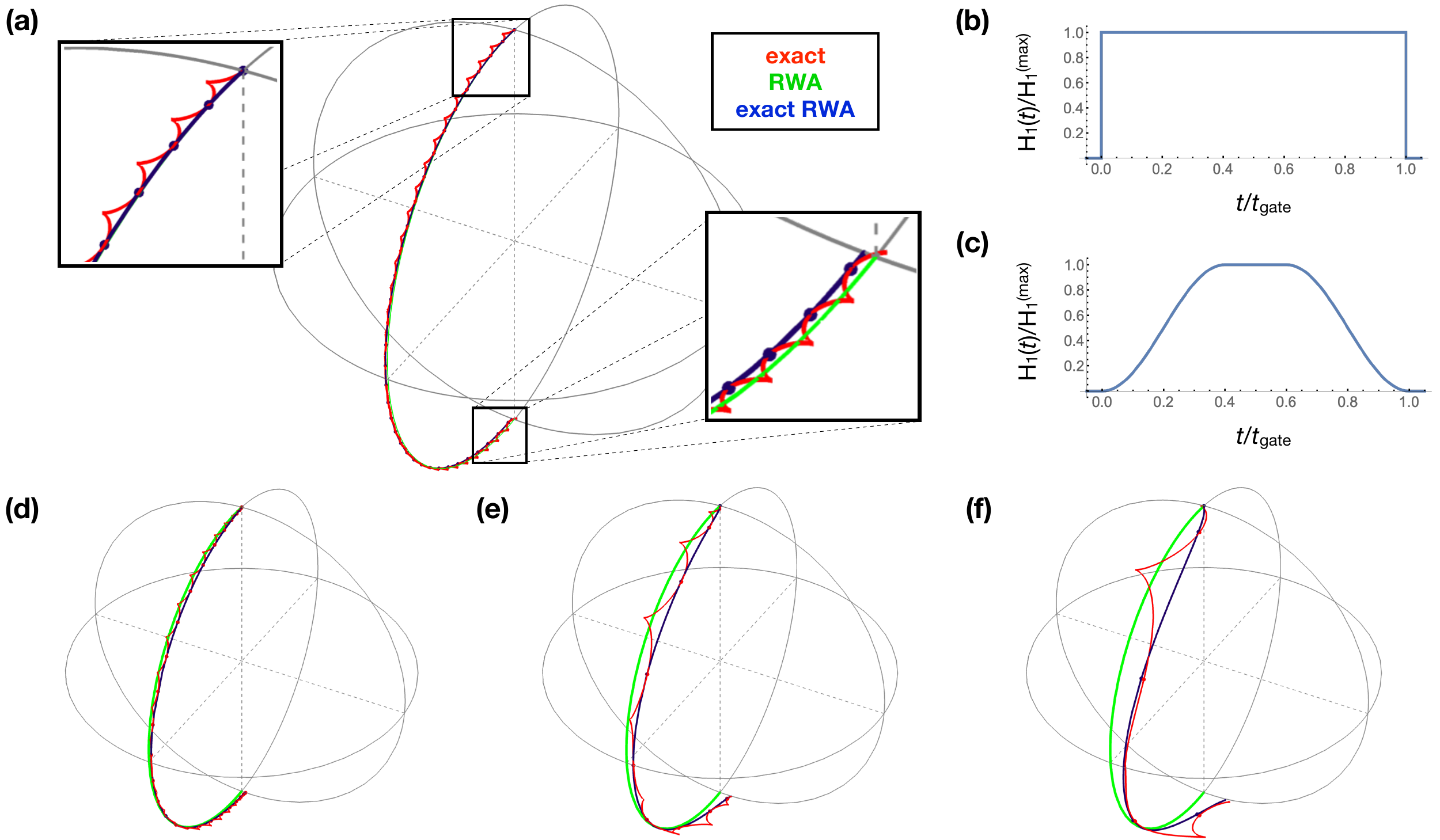}
	\caption{Various Bloch-sphere trajectories, or paths, $|\psi(t)\rangle$ [cf.~Eq.~(\ref{schroedinger})] of a driven qubit initialized to $|0\rangle$ for different envelopes corresponding to $\pi$-pulses in the RWA.  The paths shown in (a) correspond to the square pulse shown in (b) with $H_1^{(\max)}/\omega = 0.06$, while those in (d), (e) and (f) correspond to the envelope with sinusoidal ramp (defined in Note \cite{envelopeFunction}) shown in (c) for values of $H_1^{(\max)}/\omega = 0.12$, 0.33 and 0.67, respectively.  The three different paths of (a) and (d)-(f) are due to the exact (red, cycloidal-like motion), the RWA (green, no bullets) and effective Hamiltonians (blue with bullets, labeled ``exact RWA'') [see Eqs.~(\ref{Hww-2}) and (\ref{Hww-3})].  The bullets indicate the stroboscopic times (\ref{strobo}) for $t_0=0$, at which effective and exact paths coincide approximately.  In Sec.~\ref{intro:gauge_kicks} we discuss and resolve the problem that in (a) the endpoints of the effective and exact paths disagree considerably.}
	\label{trajectories}
\end{figure}

Figure \ref{trajectories} portrays the utility of the exact rotating wave approximation by means of various time evolutions of a driven qubit corresponding to an on-resonant drive with zero phase offset ($\Delta=0$ and $\phi=0$).  Here, and for all other figures in this paper, the driven-qubit trajectories are plotted in the rotating frame.  The rotating-frame Hamiltonian governing this system is given by Eq.~(\ref{Hrot2}).  In the RWA, all time evolutions in this figure correspond to single-qubit \textsc{not} gates, or $\pi$-pulses.  For the motivation of our analysis we focus on \textsc{not} gates, because in quantum information processing the \textsc{not} gate generally corresponds to the maximal manipulation applied to single qubits.  The exact Bloch-sphere trajectories $|\psi(t)\rangle$ are the solutions to the Schroedinger equation (\ref{schroedinger}), and thus follow from the time evolution operator (\ref{Ugeneric}) with $t_i = 0$ and the chosen initial condition $|\psi(t_i=0)\rangle=|0\rangle$.  The shown trajectories, or paths, then correspond to the curve traced out by the tip of the Bloch vector $|\psi(t)\rangle$.  

We contrast the exact qubit trajectories to those generated by both the RWA and effective Hamiltonians, which correspond to $|\psi(t)\rangle$ in Eq.~(\ref{schroedinger}) [with the same initial condition $|\psi(0)\rangle = |0\rangle$] upon replacing $\H_\rot$ by $\H_\RWA$ and $\H_\eff$, respectively.  Starting from the time evolution operator (\ref{Ugeneric}), the RWA trajectory is then obtained via
\begin{equation}
	\qquad \qquad U_{\RWA}(t_f, 0) = \mathcal{T}e^{-i \int_{0}^{t_f} \d \tau \mathcal \H_{\RWA}(\tau)} \stackrel{(\ref{HRWA})}{=} e^{-i \varphi \sigma_x}, \qquad\qquad\qquad (\Delta = 0, \ \phi = 0),
	\label{URWA}
\end{equation}
with $\varphi = \int_{0}^{t_f} \d \tau H_1(\tau)/4$.  In this calculation we can ignore the time ordering operator $\mathcal{T}$, because the on-resonant RWA Hamiltonian (\ref{HRWA}) commutes with itself at all times, $[\H_{\RWA}(t), \H_{\RWA}(t')]=0 \ \ \forall\ t, t'$ ($\Delta = 0$).  The condition that this pulse result in a \textsc{not} gate in the RWA, $U_{\RWA}(t_f, 0) = \mathbb{1} \cos(\varphi) + i \sigma_x \sin(\varphi) \stackrel{!}{\propto} \sigma_x$, implies $\varphi = \pi/2$, or
\begin{equation}
	\int_{0}^{t_f} \text{d}\tau\, H_1(\tau) = 2\pi.
	\label{NOT}
\end{equation}
Given that the chosen initial condition for all trajectories shown in Fig.~\ref{trajectories} is the state vector that points to the north pole of the Bloch sphere, for these $\pi$-pulses the RWA trajectories are half circles from the north to the south pole.  

For the effective time evolution operator, which also follows from Eq.~(\ref{Ugeneric}), we introduce an extended notation,
\begin{equation}
	U_{t_0}(t_f, t_0 + m t_c) = \mathcal{T} e^{-i\int{t_0+m t_c}^{t_f} \d \tau \mathcal H_{\eff}(\tau; t_0)},
	\label{Ueff}
\end{equation}
where $m$ is an integer.  In the notation introduced in Eq.~(\ref{Ueff}), the subscript $t_0$ determines the set of stroboscopic times at which the effective and exact time evolutions agree, as given in Eq.~(\ref{strobo}).  The effective trajectories shown in Fig.~\ref{trajectories} are then obtained via Eq.~(\ref{Ueff}) for the case of $t_0 = 0$ and $m = 0$.

%$\H_{\eff}$ $\H_{\rot}$
Let us compare the costs of a numerical evaluation of the various time evolution operators given above for time-dependent drive envelopes $H_1(t)$.  As noted above, due to its relatively slow time dependence the effective Hamiltonian  is significantly easier to integrate than the exact Hamiltonian.  However, we have seen that the RWA time evolution operator (\ref{URWA}) is even simpler to integrate on account of $\H_{\RWA}$ [cf.~Eq.~(\ref{HRWA})] being self-commutative at arbitrary different times.  Note that, however, for the case of off-resonant driving ($\Delta \neq 0$) the applicable RWA Hamiltonian, given in Eq.~(\ref{HRWAgeneric}) , does not generally commute with itself.  Time ordering then needs to be taken into account when computing the RWA time evolution operator, so in this case the costs of evaluating the time evolutions in the regular and exact RWA are on an equal footing.

A detailed comparison of the three different types of Bloch-sphere trajectories is given in Fig.~\ref{trajectories}(a), where the evolution's beginning and end are shown close-up.  Here the qubit is driven by a square pulse with the constant envelope $H_1^{(\max)}/\omega = 0.06$ shown in Fig.~\ref{trajectories}(b).  The exact path in Fig.~\ref{trajectories}(a), shown in red, is distinguished by its cycloidal-like motions known as Bloch-Siegert oscillations (we note that the phenomenon of the nontrivial drive dynamics in relation to these high-frequency oscillations has been commented on in Ref.~\cite{rao16}).  This is contrasted by the smooth paths in the RWA (shown in green) and in our exact RWA (blue).  For this square pulse, the Hamiltonian in either of these approximations is time-independent, thus greatly simplifying the computation of the time evolution.  Using the on-resonant RWA Hamiltonian (\ref{HRWA}), the RWA time evolution operator (\ref{URWA}) reduces to $U_{\RWA}(t_f, 0) = e^{-i H_1^{(\max)} t_f \sigma_x/4}$, resulting in a uniform circular rotation about the $x$-axis of the Bloch sphere.  Similarly, the effective time evolution operator (\ref{Ueff}) reduces to $U_{t_0=0}(t_f, m t_c) = e^{-i\H_{\eff}(t_0=0)(t_f - m t_c)}$.  Since the Hamiltonian $\H_{\eff}(t_0=0)$, given in Eq.~(\ref{Hww-2}), is a linear combination of $\sigma_x$ and $\sigma_z$, the effective trajectory corresponds to a uniform rotation about an axis in the $xz$ plane.

For a constant drive envelope, the largest deviation between the Hamiltonians in the RWA and the exact RWA is due to the Bloch-Siegert shift, the leading correction in the effective Hamiltonian (\ref{Hww-2}).  As is evident from the upper left inset in Fig.~\ref{trajectories}(a), for the sufficiently weak choice of $H_1^{(\max)}/\omega=0.06$ the paths in these two approximations are nearly indistinguishable for the first few Bloch-Siegert oscillations.  In fact, the effect of the Bloch-Siegert shift becomes noticeable only towards the end of the pulse.  The main difference between the two is that only the effective path agrees with the exact path at regular points along the entire trajectory.  These points, indicated by bullets in Fig.~\ref{trajectories}(a), correspond to the stroboscopic set of times (\ref{strobo}) for the choice of $t_0=0$.

Let us now formalize the stroboscopic time evolution that we use to define our effective Hamiltonian.  Motivated by the Bloch sphere trajectories shown in Fig.~\ref{trajectories}(a), in which the effective and exact paths agree once per Bloch-Siegert oscillation, we define the stroboscopic time evolution based on the generic time evolution operator (\ref{Ugeneric}) with initial and final times $t_i=t_0\in[0, t_c)$ and $t_f = t_0 + n t_c$,
\begin{equation}
	U_{t_0}(t_0 + n t_c, t_0) = \mathcal{T}e^{-i\int_{t_0}^{t_0 + n t_c} \d \tau \H_{\eff}(\tau; t_0)} %  = e^{-i \H_{\eff}(t_0) n t_c}
			\stackrel{!}{=} \mathcal{T}e^{-i\int_{t_0}^{t_0 + n t_c} \d \tau \H_{\rot}(\tau)} 	\qquad \forall n\in\mathbb{Z}.
	\label{goal0}
\end{equation}
Note that for the effective time evolution operator we use the same notation as above in Eq.~(\ref{Ueff}), where the subscript $t_0$ indicates the set of points (\ref{strobo}).  This equality (\ref{goal0}) of the effective and exact stroboscopic time evolution operators is a key idea in our work:  in Sec.~\ref{effectiveH} it is the point of departure for the derivation of our effective Hamiltonian defined as the series (\ref{Heff1}).  

Various qubit trajectories corresponding to the envelope shown in Fig.~\ref{trajectories}(c), which can be viewed as a smoothed square pulse, for different values of $H_1^{(\max)}/\omega$ are shown in Figs.~\ref{trajectories}(d)-(f).  Since this envelope $H_1(t)$ varies in time, the angle traversed by the exact state vector on the Bloch sphere during a Bloch-Siegert oscillation is not always the same for different segments of the trajectory.  Also, as opposed to Fig.~\ref{trajectories}(a), here the effective qubit trajectories are not simple rotations about a fixed axis, but instead follow the exact trajectory on moderately curved paths.  As in Fig.~\ref{trajectories}(a), points of agreement are marked by bullets.  For the computation of these effective Bloch sphere trajectories we have used the time evolution operator (\ref{Ueff}) with the effective Hamiltonian (\ref{Hww-3}), whose series terminates at second order in $1/\omega$.  For these example cases, we find a noticeable discrepancy between effective and exact points of (intended) agreement only in Fig.~\ref{trajectories}(f), where $H_1^{(\max)}/\omega = 0.67$.  On the contrary, in all cases of Figs.~\ref{trajectories}(a) and (d)-(f) the RWA trajectory falls short of constituting a systematic approximation to the exact trajectory.

It is the relative smoothness of the effective qubit trajectories shown in Fig.~\ref{trajectories} that makes our effective Hamiltonian theory an appealing tool for analyzing complex and perhaps unintuitive pulse shapes.  In Ref.~\cite{varvelis19}, the formalism developed in this work has been implemented and advanced in an attempt to interpret a set of highly tortuous rotations resulting in simple single-qubit gates for singlet-triplet qubits \cite{cerfontaine14}.  

The above introduction of the exact rotating wave approximation is formulated under the assumption that we are provided a smooth envelope $H_1(t)$, as stated in our assumption (\ref{assumption}).  Note that, for instance, the Hamiltonian (\ref{Hww-3}) is well defined only if the derivatives $\dot H_1$ and $\ddot H_1$ are non-singular functions.  Next, after introducing a gauge degree of freedom we consider the case of the envelope not being completely smooth.\footnote{In the present work, the attribute ``completely smooth'' is used to indicate that the function is a member of the $C^\infty$ class, i.e., its derivatives exist to all orders.}

\subsection{Gauge Freedom and Kick Operators}
\label{intro:gauge_kicks}

As described above, the set (\ref{strobo}) of stroboscopically-defined times at which the effective and exact qubit trajectories agree, is $\{t_0, t_0 \pm t_c, t_0 \pm 2t_c, \ldots \}$ with the period of the drive $t_c=\pi/\omega$ [cf.~Eq.~(\ref{tc})].  The choice of the offset time $t_0\in[0,t_c)$, a parameter that our effective Hamiltonian $\H_{\eff}(t; t_0)$ depends on, leads us to introduce a dimensionless gauge parameter,
\begin{equation}
	\qquad \qquad \qquad \qquad \beta_0 := 2\omega t_0, \qquad \qquad \qquad \qquad \qquad \beta_0 \in [0, 2\pi).
	\label{beta0}
\end{equation}
For the two effective Hamiltonians given in Eqs.~(\ref{Hww-2}) and (\ref{Hww-3}) we have chosen the specific value for the parameter $t_0=0$, thereby fixing the value of $\beta_0 = 0$.  The exemplary effective Hamiltonians in \ref{appendix:examples} are given for a variable gauge parameter $\beta_0$.  

\begin{figure}
	\includegraphics[width = \columnwidth]{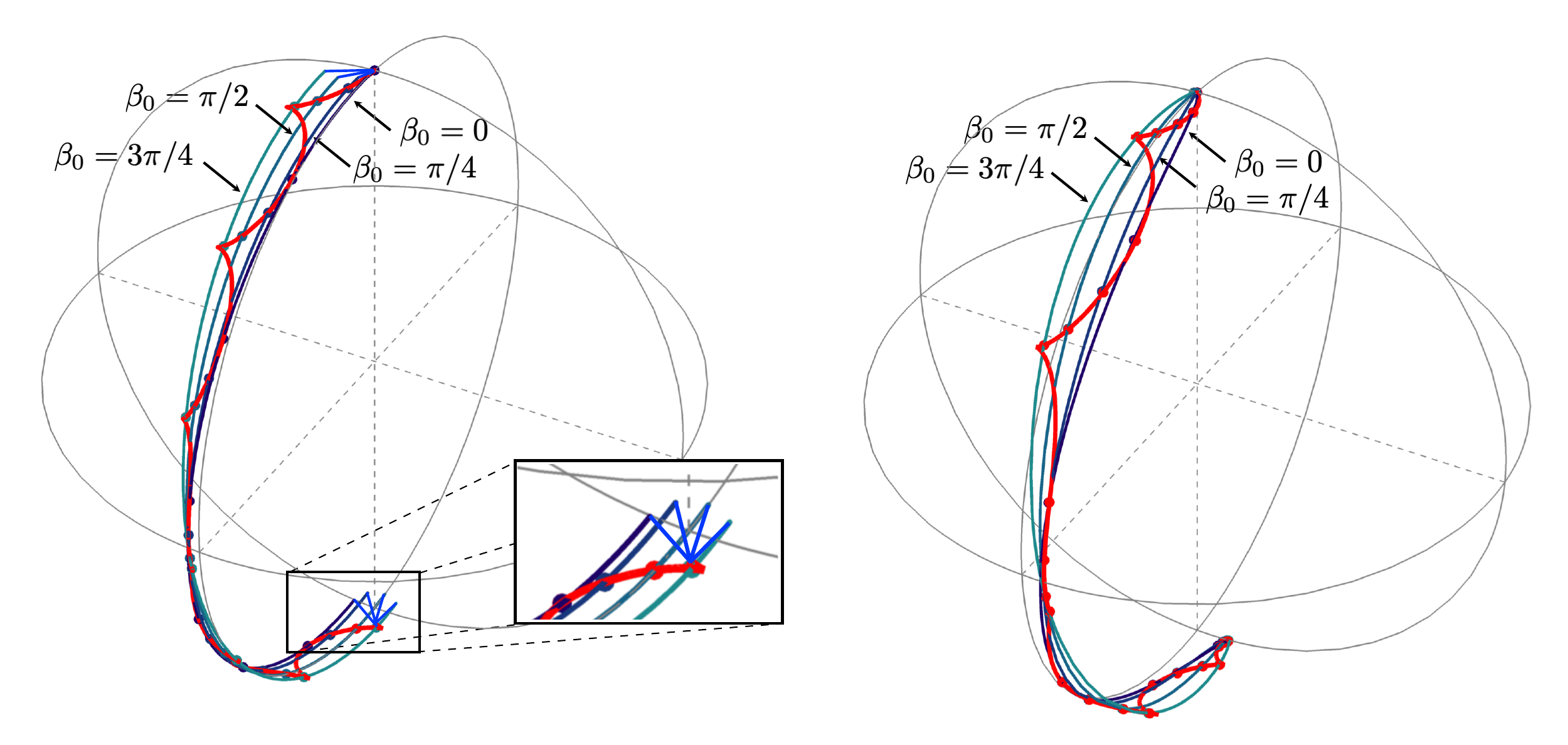}
	\caption{Various trajectories of a driven qubit initialized to $|0\rangle$.  Similar to Fig.~\ref{trajectories}, exact trajectories (shown in red) are distinguished by their cycloidal-like motions, while effective trajectories---shown for different values of the gauge parameter $\beta_0$ as indicated---appear significantly smoother.  The trajectories shown in the left and right, respectively, correspond to the envelopes shown in Figs.~\ref{trajectories}(b) and (c), each with $H_1^{(\max)}/\omega = 0.5$.  The solid straight lines on the left, appearing at the beginning and end of the paths, indicate the action of kick operators as explained in the text.}
	\label{gauge_and_kicks}
\end{figure}

The reason $\beta_0$ is called a gauge parameter is that changing its value leaves the start and end points of the effective qubit trajectory invariant.  Figure \ref{gauge_and_kicks} portrays this fact by examples of effective Bloch sphere trajectories, or paths, similar to those shown in Fig.~\ref{trajectories}, for which $t_0=0$ corresponding to $\beta_0=0$, but for various gauge parameters $\beta_0$.  As indicated in the figure, the shown effective paths are taken for the cases $\beta_0 = 0$, $\pi/4$, $\pi/2$ and $3\pi/4$.  Again, points of agreement between the effective and exact paths are indicated by bullets.  In the Bloch sphere plot on the right of Fig.~\ref{gauge_and_kicks}, which corresponds to the envelope shown in Fig.~\ref{trajectories}(c), all paths agree at the beginning, and approximately agree at the end of the pulse.  

To get a better intuition for the time evolution due to our effective Hamiltonian, we refer the interested reader to an animation of a set of Bloch sphere trajectories\footnote{\url{https://github.com/zeuch/exactRWA/tree/master/videos}\label{animation}} similar to those shown in the right of Fig.~\ref{gauge_and_kicks}, but including the RWA trajectory similar to the Bloch sphere plots shown in Fig.~\ref{trajectories}.  The piecewise-smooth envelope used for this animation is of the same functional form as that in Fig.~\ref{trajectories}(c), i.e.~it is given by the function given in Note \cite{envelopeFunction} for the case of $a=0.3$, and its maximal value is $H_1^{(\max)}/\omega = 3/14 \approx 0.21$.  The animation reveals the relative evenness of the effective motion when compared to the exact trajectory, and emphasizes that all effective trajectories (for various values of $\beta_0 \in [0, \pi]$) share the same start and end points.  

Realistic envelopes are usually not completely smooth.  Consider, for example, the two envelopes shown in Figs.~\ref{trajectories}(b) and (c).  At the beginning and end of the square pulse [see Fig.~\ref{trajectories}(b)] the envelope is discontinuous, meaning that its first derivative $\dot H_1$ is proportional to a $\delta$-function.  Similarly, the envelope with sinusoidal ramping shown in Fig.~\ref{trajectories}(c), which is defined using piecewise-analytic functions, features several divergences at its third derivative.  As a result, for both of these cases our effective Hamiltonians are not well defined at certain points in time.  This is a consequence of the assumption (\ref{assumption}) being violated if a $k$th derivative, $\dersub{H_1}{k}(t_d)$, diverges at some time $t_d$ in the form of a $\delta$-function [which corresponds to a discontinuity of the $k-1$st derivative $\dersubLong{H_1}{k-1}(t_d)$].  This problem is resolved in Sec.~\ref{nonanalytic}, where we derive kick operators to augment our formalism to deal with the following situation: for all times $t$ either Eq.~(\ref{assumption}) is fulfilled, or there exists an upper bound $l \in \mathbb{N}_0$ such that for certain times $t_d$, separated by at least the drive period $t_c$, we have
\begin{equation}
	|\dersub{H_1}{k}(t_d)| \lesssim \omega^{k+1} \quad \forall k < l, \qquad  |\dersub{H_1}{l}(t)| \propto \delta(t-t_d) + \text{nonsing}(t) \quad \forall t\in(t_d-t_c, t_d+t_c).
	\label{assumption2}
\end{equation}
Here, $\text{nonsing}(t)$ contains non-singular contributions assumed to satisfy $|\text{nonsing}(t)|\lesssim \omega^{l+1}$.  This generalized assumption (\ref{assumption2}) thus allows for envelopes like those shown in Figs.~\ref{trajectories}(b) and (c), whose derivatives violate Eq.~(\ref{assumption}) in the form of a $\delta$-function.  Section \ref{nonanalytic} determines kick operators, which cause instantaneous displacements of the effective trajectories at the times $t_d$.

For a drive whose envelope function is not smooth at $J$-many times $t_d^{(j)}$, we denote the $j$-th kick operator as $K_j(t_d^{(j)}; t_0)$.  The effective Hamiltonian then reads
\begin{equation}
	\H_\eff(t; t_0) = \H_\eff^{\text{(smooth)}}(t; t_0) + \sum_{j=1}^J K_j(t_d^{(j)}; t_0) \delta(t - t_d^{(j)}).
	\label{Heff_generic}
\end{equation}
Here, $\H_\eff^{\text{(smooth)}}$ contains only the smooth part of the Hamiltonian.  The effective time evolution operator is then a product of regular time evolutions due to $\H_\eff^{\text{(smooth)}}(t; t_0)$ and instantaneous displacements at times $t_d^{(j)}$, such as those indicated by the straight blue lines on the LHS of Fig.~\ref{gauge_and_kicks}.  An example of this situation with $J=3$ non-smooth contributions to the Hamiltonian is discussed in Sec.~\ref{nonanalytic}; this discussion includes the explicit formula of the time evolution operator [see Eq.~(\ref{U_with_kicks})].  

The effect of kick operators is illustrated by the Bloch-sphere plot on the left of Fig.~\ref{gauge_and_kicks} on the basis of the square pulse shown in Fig.~\ref{trajectories}(b).  The solid straight lines indicate instantaneous displacements in the effective qubit trajectories, caused by kick operators, at their beginning ($t_d=0$) and end ($t_d=t_{\gate}$), which is where the first derivative of the square pulse envelope is proportional to a $\delta$-function.  As shown in the figure, these displacements ensure the agreement of the exact trajectory with all effective trajectories (for different gauge parameters $\beta_0$), both at the beginning and the end of the pulse.  

Under certain circumstances, the divergence of a derivative $\dersub{H_1}{K}(t_d)$ in form of a $\delta$-function does not result in an instantaneous displacement.  As discussed in Sec.~\ref{nonanalytic}, this is the case whenever the time of the discontinuity, $t_d$, coincides with a point of stroboscopic agreement, $t_0 + n t_c$ where $t_0 = \beta_0/2\omega$ and $n\in\mathbb{Z}$.  For example, one can see no such displacement at the beginning of the $\beta_0=0$ trajectory shown in the left of Fig.~\ref{gauge_and_kicks}; this is because the first derivative of the square pulse diverges at time $t_d = 0$, which coincides with the first point of agreement at $t_0=0$.  Similarly, in the $\beta_0=0$ trajectory shown in Fig.~\ref{trajectories}(a), which also corresponds to the square pulse, there is no displacement at $t=0$.  There is, however, a displacement [not shown in Fig.~\ref{trajectories}(a)] at the end of that trajectory, which makes sure that the effective and exact trajectories agree at the very end of the pulse.  Finally, we note that the effective trajectories shown in the right of Fig.~\ref{gauge_and_kicks} also feature such displacements on account of the envelope with sinusoidal ramp having a discontinuous second derivative.  However, for the choice of drive parameters corresponding to this set of effective trajectories these displacements are too small to be noticeable in the shown Bloch sphere plot.

\section{Derivation of Effective Hamiltonian}
\label{effectiveH}

In this section we derive a recurrence relation for calculating our effective Hamiltonian, $\H_{\eff}(t; t_0)$.  As described in the Introduction, the effective Hamiltonian changes slower in time than the exact Hamiltonian, $\H_{\rot}(t)$, and generates a time evolution that agrees with the exact evolution stroboscopically.  For this derivation we assume that the amplitude function, or envelope, of the drive is completely smooth and changes slowly in time consistent with Eq.~(\ref{assumption}).  

We first show in Sec.~\ref{M} how the condition of stroboscopic agreement can be formalized naturally using the Magnus expansion.  By doing this we find that the effective Hamiltonian is determined by appropriately integrating out the fast oscillations of the drive, which are on the time scale of $1/\omega$.  However, the Magnus expansion by itself does not constitute an algebraically useful method as long as the amplitude function of the drive, $H_1(t)$, is unspecified.  Section \ref{MT} then combines the Magnus expansion with a Taylor expansion of the generic envelope $H_1(t)$, a crucial step that allows us to integrate out the fast time dependence of the exact Hamiltonian $\H_\rot(t)$ in a systematic fashion.  We refer to this combined method as the Magnus-Taylor expansion.  In Sec.~\ref{procedure} we apply this Magnus-Taylor expansion to the driven-qubit problem, thereby rendering the condition of stroboscopic agreement algebraically tractable.  We then solve this condition asymptotically to arrive at the central result of our work:  a recurrence relation for computing the effective Hamiltonian series (\ref{Heff1}).

Subsequently, in Sec.~\ref{HeffExample} we exemplify the usage of our central recurrence relation by computing an effective Hamiltonian up to first order in $1/\omega$.  Section \ref{Heffconst} closes our treatment of smooth amplitude functions by deriving a simplified computation method for determining the effective Hamiltonian of the form (\ref{Heff0}) for the special case of a constant drive envelope.

\subsection{Applying the Magnus Expansion}
\label{M}

Our derivation of the effective Hamiltonian employs the Magnus expansion \cite{magnus1954, ernst87, waugh07}, which is a variant of time-dependent perturbation theory carried out at the level of the Hamiltonian rather than the wave function.  A central characteristic of this method is that, independent of the order at which this expansion series is truncated, its implementation inherently conserves unitarity of the time evolution.  

The basic idea behind this expansion is to write the time evolution operator as a true exponential function of a Magnus expansion $\overline\H$, which is calculated perturbatively.  Assuming a generic Hamiltonian $\H(t)$, we rewrite the time evolution operator (\ref{Ugeneric}) using the Magnus expansion as follows,
\begin{equation}
	U(t_f, t_i) = \mathcal{T}e^{-i\int_{t_i}^{t_f} \d \tau \mathcal H(\tau)} = e^{-i \overline{\mathcal H} (t_f - t_i)}. %\quad
	\label{UM1}
\end{equation}
A straightforward way to obtain a series representation of $\overline{\mathcal H}$ is to expand the exponential functions on both sides of Eq.~(\ref{UM1}), apply the time ordering operator and equate terms of equal power in $\lambda = -i$ \cite{waugh07}.  We note that while the Magnus expansion depends on the initial and final times $t_i$ and $t_f$, respectively, we avoid uncomfortable overloading in the notation introduced in Eq.~(\ref{UM1}) by suppressing this dependence.  

We focus on the stroboscopic time evolution for the set of times $\{t_0, t_0 \pm t_c, \ldots \}$ [cf.~Eq.~(\ref{strobo})] with the period $t_c=\pi/\omega$ of the drive in the rotating frame [cf.~Eq.~(\ref{tc})] and a time offset $t_0\in[0, t_c)$.  The stroboscopic time evolution operator similar to that in Eq.~(\ref{goal0}) can be written via the Magnus expansion as
\begin{equation}
	U_{t_0}(t_0 + n t_c, t_0) = e^{-i \overline{\mathcal H} n t_c},
	\label{UM2}
\end{equation}
for integers $n$.  The Magnus expansion $\overline \H$ can be viewed as the result of computing a ``sophisticated average'' over the interval $[t_0, t_0 + n t_c)$, which we call a \emph{Magnus interval}.  Written explicitly for such a Magnus interval, the standard procedure for obtaining this Magnus expansion is to write $\overline \H$ as a series
\begin{equation}
	\overline \H = \sum_{k=0}^\infty \overline \H^{(k)}.
	\label{MagnusSeries}
\end{equation}
The first three of the terms $\H^{(k)}$, which below we refer to as \emph{Magnus integrals}, are given by
\begin{eqnarray}
	\overline \H^{(0)} &=& \frac1{n t_c} \int_{t_0}^{t_0 + nt_c} \text d \tau \H(\tau), \label{Hb0}\\
	\overline \H^{(1)} &=& \frac{-i}{2 nt_c} \int_{t_0}^{t_0 + nt_c}\text d \tau' \int_{t_0}^{\tau'} \text d \tau [\H(\tau'),\H(\tau)], \label{Hb1}\\
	\overline \H^{(2)} &=& -\frac{1}{6 nt_c} \int_{t_0}^{t_0 + nt_c}\text d \tau'' \int_{t_0}^{\tau''} \text d \tau' \int_{t_0}^{\tau'} \text d \tau \Big\{[\H(\tau''),[\H(\tau'),\H(\tau)]] + [[\H(\tau''),\H(\tau')],\H(\tau)]\Big\}. \quad \label{Hb2}
\end{eqnarray}
Magnus integrals of higher order may be determined recursively (see, e.g., Refs.~\cite{blanes09,blanes10}) or as outlined below Eq.~(\ref{UM1}).  

\begin{figure}
	\centering
	\includegraphics[width=0.9\columnwidth]{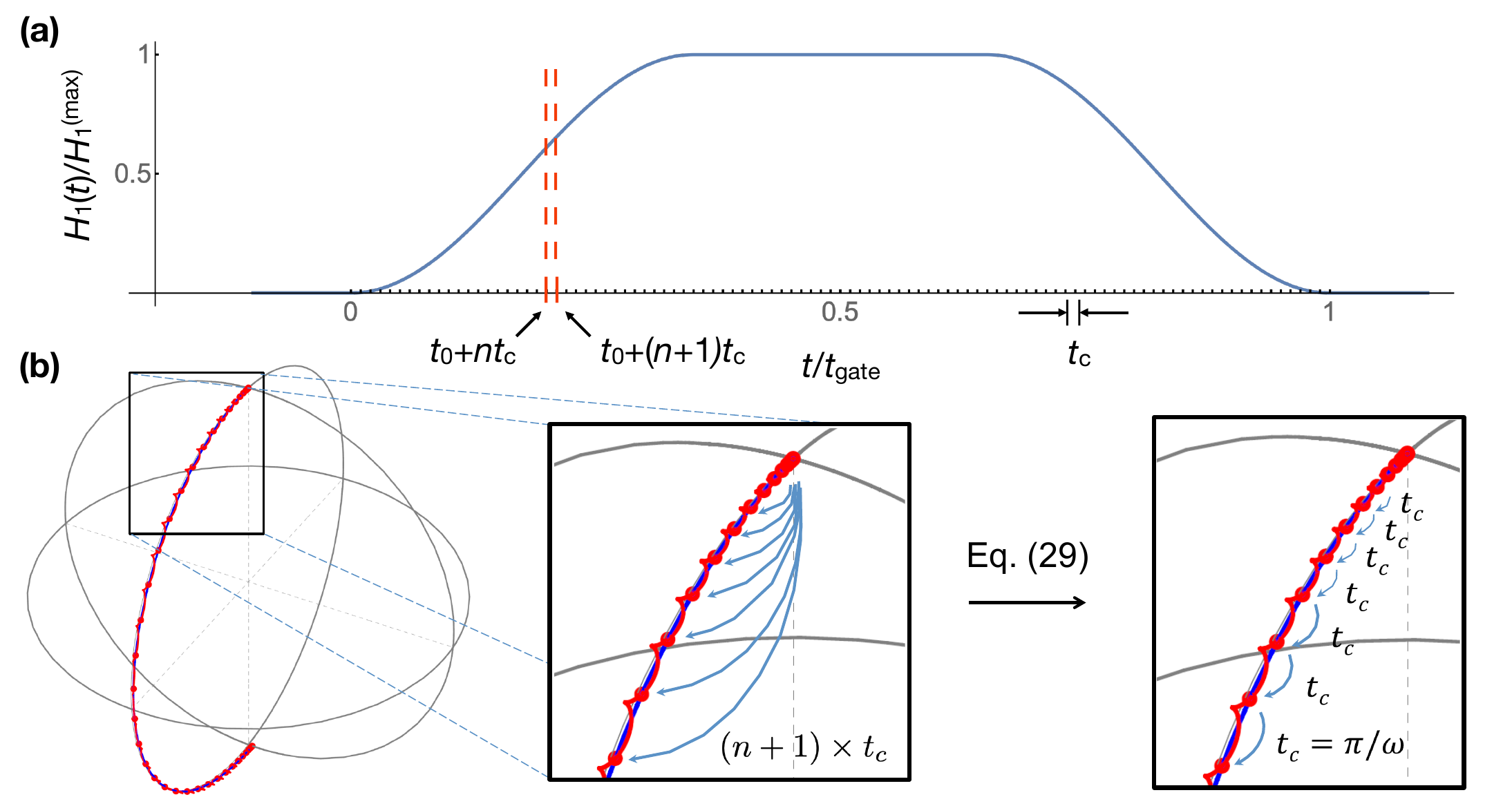}
	\caption{Graphical illustrations of several concepts used in our derivation.  (a) Generic drive envelope $H_1(t)$, where the time axis is divided into Magnus intervals of duration $t_c = \pi/\omega$.  As discussed in the text, this time $t_c$ is short compared to the gate duration $t_\gate$ for weak pulses with $H_1^{(\max)}/\omega\lesssim 0.01$.  In (b) the stroboscopic time evolution for a generic value of the gauge parameter $\beta_0$ is shown in two equivalent representations, which can be related to one another using the property (\ref{property}) of the time evolution operator.  This property allows us to simplify the condition (\ref{condition}) of the effective Hamiltonian to the condition (\ref{condition2.5}), which is based on the Magnus average on the fundamental Magnus interval (\ref{fundamental}).}
	\label{H1_distribute}
\end{figure}

Below we begin our derivation by re-expressing the condition of stroboscopic time evolution [cf.~Eq.~(\ref{goal0})] using the  Magnus expansion.  Figure \ref{H1_distribute} illustrates the central ideas that precipitate our line of arguments leading to a sound condition on our effective Hamiltonian, which can be solved asymptotically.  A generic pulse envelope $H_1(t)$ for a $\pi$-pulse, or a single-qubit \textsc{not} gate, in the rotating wave approximation is shown in Fig.~\ref{H1_distribute}(a).  As stated in Eq.~(\ref{NOT}), for such a pulse the area under the $H_1(t)$ function is $2\pi$.  Figure \ref{H1_distribute}(a) makes it evident that for relatively weak fields satisfying $H_1^{(\max)}/\omega \lesssim 0.01$ the gate duration of the $\pi$-pulse, $t_\gate$, will cover many Magnus intervals of duration $t_c = \pi/\omega$.  This suggests that for such weak fields the Magnus series (\ref{MagnusSeries}) of the exact Hamiltonian $\H_{\rot}$ is likely to converge quickly over a Magnus interval of duration $t_c$.  Based on this intuition, in the present section we seek a condition for the effective Hamiltonian formulated using the Magnus expansion for precisely one such interval of duration $t_c$.  

The basic condition of stroboscopic time evolution is expressed in Eq.~(\ref{goal0}) by the equality of the effective and exact time evolution operators at the stroboscopic times (\ref{strobo}).  We reformulate this condition (\ref{goal0}) using the Magnus expansion by, for convenience in the following analysis, shifting the final time by one period $t_c$ with respect to the time evolution operator (\ref{UM2}), $t_f \rightarrow t_0 + (n+1) t_c$,
\begin{equation}
	U_{t_0}(t_0 + (n+1) t_c, t_0) = e^{-i \overline\H_{\eff} (n+1) t_c} \stackrel{!}{=} e^{-i \overline \H_{\rot}(n+1) t_c} \quad \qquad \forall n \in \mathbb{Z}.
	\label{condition}
\end{equation}
To be clear, here each Magnus expansion is to be evaluated on the Magnus interval $[t_0, t_0 + (n+1) t_c)$.  This stroboscopic time evolution over a duration $(n+1) t_c$ is illustrated by the close-up view of the Bloch sphere plot shown in the left-hand side of Fig.~\ref{H1_distribute}(b) for the envelope shown in Fig.~\ref{H1_distribute}(a).  

Using the group property of the time evolution operator $U$ from initial time $t_0$ to final time $t_0+(n+1)t_c$,
\begin{equation}
	U(t_0 + (n+1)t_c, t_0) =\, \ \op_{j=0}^{n} \ U(t_0 + (j+1) t_c, t_0 + j t_c) \qquad \forall n \in \mathbb{Z},
	\label{property}
\end{equation}
where the symbol $\ \op \ $ signifies that the product is ordered from right to left with increasing $j$, we rewrite the conditions (\ref{condition}) as the set of conditions
\begin{equation}
	U_{t_0}(t_0 + (n+1) t_c, t_0 + n t_c) = e^{-i \overline\H_{\eff} t_c} \stackrel{!}{=} e^{-i \overline\H_{\rot}t_c} \qquad \forall n \in \mathbb{Z}.
	\label{condition2}
\end{equation}
The transition from the time evolution operator (\ref{condition}) to this stepwise time evolution is also illustrated in Fig.~\ref{H1_distribute}(b).  It is thus evident that the effective Hamiltonian is fundamentally defined via the Magnus expansion for a stroboscopic evolution over each individual Magnus interval $[t_0+ n t_c, t_0 + (n+1) t_c)$,
\begin{equation}
	\hspace{3cm} \overline\H_{\eff} \stackrel{!}{=} \overline\H_{\rot}, \quad \qquad [\text{all Magnus intervals } [t_0 + n t_c, t_0 + (n+1) t_c)].
	\label{condition2.4}
\end{equation}

We now discuss how the conditions (\ref{condition2.4}), which differ from one another through the dependence on the parameter $n$ counting the Magnus interval, can be effectively reduced to a simpler set of conditions, all of which (i) are of the same form, and (ii) require taking the Magnus average over one and the same Magnus interval.  We choose this particular interval, henceforth called the \emph{fundamental Magnus interval}, to be
\begin{equation}
	\qquad \qquad \qquad [t_0, t_0 + t_c)\qquad \qquad \qquad \,\,\,\,\,\,\,\,\,\text{(fundamental Magnus interval)}.
	\label{fundamental}
\end{equation}
Let us discuss how this reduction comes about.  Notice that the first three terms of each Magnus expansion in Eq.~(\ref{condition2.4}) can be obtained from Eqs.~(\ref{Hb0})-(\ref{Hb2}) by replacing all lower integral bounds in these equations according to the rule $t_0 \rightarrow t_0 + n t_c$, and further replacing each outermost upper integral bound according to $t_0 + n t_c \rightarrow t_0 + (n+1) t_c$.  Each such term can then be reduced to that of $n=0$, corresponding to the Magnus expansion over the fundamental Magnus interval, by shifting the integration variables and introducing a relocated envelope $\tilde H_1$ as follows,
\begin{equation}
	\qquad \qquad \tau \rightarrow \tilde \tau = \tau - n t_c,	
	\qquad \qquad 
	\tilde H_1(\tilde \tau) := H_1(\tilde \tau + n t_c) = H_1(\tau).
	\label{trafo}
\end{equation}

We exemplify this reduction by considering the lowest-order Magnus term $\overline \H^{(0)}$, given in Eq.~(\ref{Hb0}), for the rotating-frame Hamiltonian (\ref{Hrot2}) for the special case of $\Delta = 0$ and $\phi=0$, i.e., $\H_{\rot}(t) = (H_1(t)/4) ( \sigma_x + \cos(2\omega t) \sigma_x - \sin(2\omega t)\sigma_y)$.  Examining only the terms proportional to $\sigma_x$, the integral of interest can be straightforwardly manipulated using the rules (\ref{trafo}) as follows,
\begin{eqnarray}
	\frac{1}{t_c}\int_{t_0 + n t_c}^{t_0 + (n+1) t_c} \d \tau \frac{H_1(\tau)}{4} (1+\cos(2\omega \tau))\sigma_x &=& \frac{1}{t_c}\int_{t_0}^{t_0 + t_c} \d \tilde \tau \frac{H_1(\tilde \tau + n t_c)}{4} (1+\cos(2\omega (\tilde \tau+n t_c))) \sigma_x \nonumber \\
			&=& \frac{1}{t_c}\int_{t_0}^{t_0 + t_c} \d \tilde \tau \frac{\tilde H_1(\tilde \tau)}{4}  (1+\cos(2\omega \tilde \tau)) \sigma_x.
	\label{n->0}
\end{eqnarray}
The transformation of the terms proportional to $\sigma_y$ can be done in parallel.  We have thus transformed the lowest-order Magnus integral over a generic Magnus interval to an equivalent integral over the fundamental Magnus interval (\ref{fundamental}).  Similar transformations can be used to reduce other terms of the Magnus series, such as those in Eqs.~(\ref{Hb1}) and (\ref{Hb2}), to the same fundamental interval.  

Above, starting from Eq.~(\ref{condition}) we have transformed the condition on equal time evolution at the time points (\ref{strobo}) to the infinite set of conditions (\ref{condition2.4}).  These conditions differ from one another only by the Magnus interval that the Magnus expansion is based on.  Given that, as we have just discussed, each Magnus expansion can be reduced to one and the same Magnus interval, we now focus on the single condition
\begin{equation}
	\hspace{3cm} \overline\H_{\eff} \stackrel{!}{=} \overline\H_{\rot}  \quad \qquad [\text{fundamental Magnus interval } [t_0 , t_0 + t_c)].
	\label{condition2.5}
\end{equation}
We note that since we have now established that we can concentrate on the fundamental Magnus interval, below we will not continue to use the notation for $\tilde H_1$ and $\tilde \tau$ introduced in Eq.~(\ref{trafo}) by assuming $t\in[t_0 , t_0 + t_c)$.  This condition (\ref{condition2.5}) is the key property of the effective Hamiltonian that enables us to derive the desired recurrence relation in Sec.~\ref{procedure}.  

\subsection{Magnus-Taylor Expansion}
\label{MT}

Without specification of the amplitude function $H_1(t)$, closed-form expressions of the Magnus expansions that appear in the previous section cannot be obtained.  For example, consider the Magnus expansion $\overline \H_{\rot}$ in Eq.~(\ref{condition2.5}).  Focusing on the rotating-frame Hamiltonian (\ref{Hrot2}) and the lowest-order Magnus term (\ref{Hb0}) [with $n=1$ since Eq.~(\ref{condition2.5}) is constrained to the fundamental Magnus interval (\ref{fundamental})], the integral in question,
\begin{equation}
	\overline \H_{\rot}^{(0)} = \frac1{t_c} \int_{t_0}^{t_0 + t_c} \d \tau \H_{\rot}(\tau) = \frac1{t_c} \int_{t_0}^{t_0 + t_c} \d \tau \frac{H_1(\tau)}4 ( \sigma_x + \cos(2\omega \tau) \sigma_x - \sin(2\omega \tau)\sigma_y),
	\label{not-simplifiable}
\end{equation}
cannot be simplified any further.  Note that this integral \emph{can}, however, be solved as a series expansion in $1/\omega$ upon replacing $H_1(\tau)$ by its Taylor series with respect to a reference time $t$,
\begin{equation}
	H_1(\tau) = H_1(t) + \dot H_1(t) (\tau - t) + \frac 12 \ddot H_1(t) (\tau - t)^2 + \ldots,
	\label{TaylorDots}
\end{equation}
where we take this reference time $t \in [t_0, t_0 + t_c)$ so that $|\tau-t| \leq t_c \sim 1/\omega$.  Given our assumption of a slowly varying envelope $H_1(t)$ as expressed by Eq.~(\ref{assumption}), this Taylor series converges quickly for weak fields $|H_1(t)|/\omega\lesssim 0.01$ since these, referring to Fig.~\ref{H1_distribute}(a), correspond to the case of Magnus intervals that are short compared to the total duration of the pulse.  

Suppose we wanted to determine the zeroth-order Magnus term (\ref{Hb0}) for the effective Hamiltonian (\ref{Hww-3}), which depends not only on the envelope $H_1(t)$ but also on its first and second derivatives, $\dot H_1(t)$ and $\ddot H_1(t)$.  In this case, we would encounter the integral (\ref{not-simplifiable}) upon replacing $\H_{\rot}(\tau) \rightarrow \H_{\eff}(\tau; t_0)$.  To solve such an integral explicitly, our approach would then involve introducing three separate Taylor series similar to Eq.~(\ref{TaylorDots}) with one and the same reference time $t$ for all three functions $H_1(\tau)$, $\dot H_1(\tau)$ and $\ddot H_1(\tau)$.  This perturbative procedure may also be used for all higher-order Magnus terms of the series (\ref{MagnusSeries}), and it is this very strategy that constitutes the Magnus-Taylor expansion formalized below.  

We note that the above procedure of integrating out the quickly-oscillating terms of the Hamiltonian is reminiscent of the two-time Floquet formalism \cite{sambe73, howland74, breuer89ZPD, peskin93, drese99, novicenko17} in which a similar separation between slow and fast temporal dependencies takes place.  Furthermore, the most recent of these studies, Ref.~\cite{novicenko17}, presents effective Hamiltonians that similarly depend on the first derivative of slowly varying parameters of the Hamiltonian.  

Now consider a generic Hamiltonian $\H$ that may have both explicit and implicit dependence on time $t$, for which the implicit time dependence is mediated by parameters ${\bf X}(t) = \{ X_1(t), X_2(t), \ldots \}$, i.e., $\H(t) = \H(t, {\bf X}(t))$.  For instance, the rotating-frame Hamiltonian $\H_{\rot}(t)=\H_{\rot}(t, \{H_1(t)\})$ given in Eq.~(\ref{Hrot0}) has one such time-dependent parameter, such that ${\bf X}(t) = \{H_1(t)\}$.  In contrast, the example effective Hamiltonian (\ref{Hww-3}), while not explicitly time-dependent, depends on three parameters ${\bf X}(t) = \{H_1(t), \dot H_1(t), \ddot H_1(t)\}$, i.e., $\H_\eff(t)=\H_\eff(\{H_1(t), \dot H_1(t), \ddot H_1(t)\})$.  Each time-dependent parameter $X_i$ hence appearing in the Magnus integrals, such as those given in Eqs.~(\ref{Hb0})-(\ref{Hb2}), is a function of an integration variable, denoted $\tau$ below.  We denote the Taylor series of such a function $X_i(\tau)$ with respect to a reference time $t$ by
\begin{equation}
	\T[X_i(\tau), t] := \sum_{k = 0}^{\infty} \frac{\dersub{X_i}{k}(t)}{k!} (\tau-t)^k.
	\label{Taylorexpansion}
\end{equation}
We further denote the Taylor series of the vector ${\bf X}(\tau)$ with respect to the same reference time $t$ by
\begin{equation}
	\T[{\bf X}(\tau), t] = \{\T[X_1(\tau), t], \T[X_2(\tau), t], \ldots \}.
	\label{TaylorexpansionVector}
\end{equation}

We now define the Magnus-Taylor expansion for a Hamiltonian $\H(t, {\bf X}(t))$ as a regular Magnus expansion with the added feature that every function $X_i(\tau)$ appearing in the integrals of this expansion is replaced by a Taylor series.  Let $\M[\H,  t; t_0]$ denote this Magnus-Taylor expansion of a Hamiltonian $\H$, a reference time $t$ and the fundamental Magnus interval $[t_0, t_0 + t_c)$ [cf.~Eq.~(\ref{fundamental})].  The Magnus-Taylor expansion is then defined analogously to the Magnus series (\ref{MagnusSeries}) for $n=1$,
\begin{equation}
	\M[\H, t; t_0] = \sum_{k = 0}^{\infty} \m_k[\H, t; t_0].
	\label{MagnusTaylor}
\end{equation}
Here the term $\m_k$ is the $k$th-order Magnus term in which each parameter $X_i(\tau)$, which enters the Magnus integrals through the Hamiltonian $\H(\tau, {\bf X}(\tau))$, is replaced by its Taylor series $\T[X_i(\tau), t]$ with reference time $t$ as given in Eq.~(\ref{Taylorexpansion}).  For example, the first two terms in the sum of Eq.~(\ref{MagnusTaylor}) are given by, referring to Eqs.~(\ref{Hb0}) and (\ref{Hb1}) setting $n=1$,
\begin{eqnarray}
	\m_0[\H, t; t_0] &=& \frac1{t_c} \int_{t_0}^{t_0+t_c} \text d \tau \H(\tau, \T[{\bf X}(\tau), t]), \label{mt0} \\
	\m_1[\H, t; t_0] &=& \frac{-i}{2t_c} \int_{t_0}^{t_0+t_c}\text d \tau' \int_{t_0}^{\tau'} \text d \tau [\H(\tau', \T[{\bf X}(\tau'), t]), \H(\tau, \T[{\bf X}(\tau), t])]. \label{mt1}
\end{eqnarray}
As noted below Eq.~(\ref{TaylorDots}), for reasons of convergence we assume that the reference time $t$ for this Magnus-Taylor expansion lies within the fundamental Magnus interval, or $t\in[t_0, t_0 + t_c)$.  

The Magnus-Taylor expansion can be used to compute the time evolution of a quantum system in the same way as the regular Magnus expansion.  For example, the time evolution operator across the fundamental Magnus interval (\ref{fundamental}) can be expressed as
\begin{equation}
	U(t_0 + t_c, t_0) = e^{-i \M[\H, t; t_0] t_c},
	\label{UMT}
\end{equation}
which is analogous to the expression using the Magnus expansion in Eq.~(\ref{UM2}) for the case of $n=1$.  In Sec.~\ref{nonanalytic}, we use a Magnus-Taylor expansion in a similar manner as in Eq.~(\ref{UMT}) to determine the time evolution over a partial Magnus interval in when deriving the kick operators for non-smooth drive envelopes.  

Above we have reduced the set of conditions (\ref{condition2.4}) for the effective Hamiltonian $\H_{\eff}$, which features Magnus expansions $\overline \H$ on Magnus intervals $[t_0 + n t_c, t_0 + (n+1) t_c)$ determined by $n$ and $t_0$, to the single condition (\ref{condition2.5}) featuring the Magnus expansion on the \emph{fundamental} Magnus interval $[t_0, t_0 + t_c)$.  In our derivation in Sec.~\ref{procedure}, we replace each Magnus expansion in Eq.~(\ref{condition2.5}) by the more tractable Magnus-Taylor expansion $\M[\H, t; t_0]$, which requires the specification of an additional parameter, the reference time $t$.  In the same way as $t_0$ is associated with the effective time evolution via the gauge parameter, $\beta_0 = 2\omega t_0$, we associate the \emph{reference time} $t$ of the Magnus-Taylor expansion with the \emph{current time} $t$ of the effective Hamiltonian $\H_{\eff}(t; t_0)$.  

To give an example, let us calculate the Magnus-Taylor expansion of the Hamiltonian (\ref{Hrot2}) up to first order in $1/\omega$.  For this order, the only relevant terms in the Magnus-Taylor series (\ref{MagnusTaylor}) are $\m_0$ and $\m_1$,
\begin{equation}
	\M[\H_{\rot}, t; t_0] = \m_0[\H_{\rot}, t; t_0] + \m_1[\H_{\rot}, t; t_0] + \O(1/\omega^2).
	\label{MagnusTaylor1}
\end{equation}
For the first term (\ref{mt0}) we truncate the Taylor series (\ref{Taylorexpansion}) at first order,
\begin{eqnarray}
	\m_0[\H_{\rot}, t; t_0] &=& \frac 1{t_c} \int_{t_0}^{t_0+t_c} \text d \tau \left(\frac{H_1(t)}4 + \frac{\dot H_1(t)}4 (\tau-t) + \O((\tau-t)^2)\right) (\sigma_x + \cos(2\omega \tau)\sigma_x - \sin(2\omega \tau)\sigma_y) \nonumber \\
			&=& \frac{H_1(t)}4 \sigma_x + \frac{\dot H_1(t)}{8 \omega} ((\omega t_c + 2\omega(t_0-t) + \sin(2\omega t_0)) \sigma_x + \cos(2\omega t_0) \sigma_y) + \O(1/\omega^2)  \nonumber\\
			&\stackrel{(\ref{HRWA})}{=}& H_{\RWA}(t) + \frac{\dot H_1(t)}{8 \omega} ((\pi + \beta_0 - \beta + \sin \beta_0) \sigma_x + \cos \beta_0 \sigma_y)+ \O(1/\omega^2).
	\label{mt0_example}
\end{eqnarray}
In the last line we took note that in the Magnus-Taylor expansion the lowest order in $1/\omega$ results in the Hamiltonian in the rotating wave approximation (RWA), given by Eq.~(\ref{HRWA}).  We also replaced $t_0$ using $\beta_0 = 2\omega t_0$ as defined in Eq.~(\ref{beta0}), and introduced the dimensionless quantity
%In the last line we took note that in the Magnus-Taylor expansion the lowest order in $1/\omega$ results in the RWA Hamiltonian (\ref{HRWA}) and, in analogy to $\beta_0 = 2\omega t_0$ in Eq.~(\ref{beta0}), we introduced
\begin{equation}
	\beta := 2\omega t
	\label{beta}
\end{equation}
in analogy to $\beta_0$.  This, together with using $t_c = \pi/\omega$, serves the purpose of consistently separating different orders of $1/\omega$ within the expansion.  

To compute the second term (\ref{mt1}) of the Magnus-Taylor expansion, note that for $\H_{\rot}$ as given in Eq.~(\ref{Hrot2}), the commutator $[\H_{\rot}(\tau), \H_{\rot}(\tau')]$ is proportional to $\sigma_z$.  Here, we truncate the Taylor series (\ref{Taylorexpansion}) already at zeroth order,
% % note:  [\H, \H] ~ (1/4)^2 * {[1+cos(2x)]sin(2y) - sin(2x)[1+cos(2y)]} = -(1/4)^2*4 cos(x) cos(y) sin(x - y) [according to wolframalpha.com]
\begin{eqnarray}
	\m_1[\H_{\rot}, t; t_0] &=& \frac {-i}{2t_c} \int_{t_0}^{t_0+t_c} \text d \tau' \int_{t_0}^{\tau} \d\tau \frac 14 (H_1(t)+ \O(\tau'-t)) (H_1(t)+ \O(\tau-t)) \cos(\omega \tau') \nonumber \\ 
			&& \qquad \qquad \qquad \qquad\qquad\qquad \qquad \qquad\qquad\qquad \times \cos(\omega \tau) \sin(\omega (\tau-\tau'))[\sigma_x, \sigma_y] \nonumber \\
			&=& \frac {1}{t_c} \int_{t_0}^{t_0+t_c} \text d \tau' \frac{H_1(t)^2}{16} \cos(\omega \tau')( \cos(\omega \tau') - \cos(2\omega \tau')) \sigma_z + \O(1/\omega^2) \nonumber \\
			&=& \frac{H_1(t)^2}{32\omega} (1-2\cos (2\omega t_0)) \sigma_z + \O(1/\omega^2) \nonumber\\
			&=& \frac{H_1(t)^2}{32\omega} (1-2\cos\beta_0) \sigma_z + \O(1/\omega^2),
	\label{mt1_example}
\end{eqnarray}
where, as in Eq.~(\ref{mt0_example}), we express the result in terms of $\beta_0$.

As in the above example, in the remainder of this work we express all temporal parameters that have an impact on the dimension of the Magnus-Taylor expansion using dimensionless quantities.  As a result, the only dimension\emph{ful} quantities in our expressions, besides the drive frequency $\omega$, are the detuning, $\Delta$, the field strength $H_1(t)$ and its time derivatives $\dersub{H_1}{k}(t)$.  Hence, in the present study the coefficients of the operators $\sigma_x$, $\sigma_y$ and $\sigma_z$ in a Magnus-Taylor term of order $1/\omega^k$, or simply of order $k$, are of the form
\begin{equation}
	\frac{\Delta^{n_{0}} H_1^{n_1} \dot H_1^{n_2} \ddot H_1^{n_3} \ldots \Big(\dersub{H_1}{k}\Big)^{n_{k+1}} }{\omega^{k}},
	\label{coefficients}
\end{equation}
which is given up to a dimensionless factor.  Note that the Magnus-Taylor expansion yields an operator that has the same units as a Hamiltonian, because of which the exponents $n_i$ with $i = 0, 1, \ldots, k+1$ can be found to fulfill the requirement $n_{0}  + \sum_{j=1}^{k+1} j n_j = k + 1$.  By construction, for any Magnus-Taylor term that appears in this study the only dimensionful quantity in the denominator of the coefficient (\ref{coefficients}) is $\omega$, corresponding to non-negative integers $n_i$.  Examples of such coefficients for terms of orders $k=0$ or 1 can be found above in Eqs.~(\ref{mt0_example}) and (\ref{mt1_example}).  In these calculations we computed only the first two terms $\m_0$ and $\m_1$ of the series (\ref{Taylorexpansion}), and we kept at most the first two terms in the Taylor expansion (\ref{MagnusTaylor}).  

In practice, these series may always be truncated at an appropriate order.  For a Magnus-Taylor term of order $k$ with coefficient (\ref{coefficients}), the largest possible exponent $\max(n_0, n_1, \ldots)$ is directly related to the highest depth of the commutator of the term $\m_k$ in Eq.~(\ref{MagnusTaylor}) [for $\m_0$ and $\m_1$, for instance, see Eqs.~(\ref{mt0}) and (\ref{mt1})], and its value coincides with the highest required term in the Magnus-Taylor series (\ref{MagnusTaylor}).  Similarly, the highest temporal derivative coincides with the highest required term in the Taylor series (\ref{Taylorexpansion}).  For lowest order in the Magnus expansion, $k=0$, the only terms permitted are those with coefficients (\ref{coefficients}) equal to $\Delta$ or $H_1$, i.e., the only nonzero exponents are $n_0=1$ or $n_1=1$.  This implies that in both Eqs.~(\ref{Taylorexpansion}) and (\ref{MagnusTaylor}) only the $k=0$ terms need to be kept.  For the next order of $k=1$, or $1/\omega$, the highest required terms in the same two equations are those with $k=1$, since here the possible coefficients (\ref{coefficients}) are given by $\Delta H_1/\omega$, $H_1^2/\omega$ and $\dot H_1/\omega$.  The same dimensional argument can be applied to arbitrary orders, with the corresponding result that for a Magnus-Taylor expansion of order $k$ both Eqs.~(\ref{Taylorexpansion}) and (\ref{MagnusTaylor}) can be truncated at order $k$. 

In this work, the purpose of the Magnus-Taylor expansion is twofold.  First, in the following Sec.~\ref{procedure} we use this expansion directly to derive our effective Hamiltonian.  Second, in Sec.~\ref{nonanalytic} we use it to compute various time evolution operators similar to that in Eq.~(\ref{UMT}).  This allows us to derive kick operators, which extend our effective-Hamiltonian theory for amplitude functions that are not entirely smooth.

\subsection{Recurrence Relation for Effective Hamiltonian}
\label{procedure}

We are now in a position to derive an explicit condition that enables us to find the recurrence relation for constructing our effective Hamiltonian (\ref{Heff1}).  We start by rewriting the previous condition (\ref{condition2.5}) using the Magnus-Taylor expansion (\ref{MagnusTaylor}),
\begin{equation}
	\M[\H_{\eff}, t; t_0] \stackrel{!}{=} \M[\H_{\rot}, t; t_0].
	\label{condition3}
\end{equation}
As discussed in the previous section, the reference time $t$ of this Magnus-Taylor expansion corresponds to the current time $t$ of the effective Hamiltonian, $\H_\eff(t; t_0)$, and we assume $t\in[t_0, t_0+t_c)$.  The goal expressed by Eq.~(\ref{condition3}) is to obtain an effective Hamiltonian whose Magnus-Taylor expansion is equal to that of the rotating-frame Hamiltonian.  

Denoting the $k$th coefficient of a power series, $p(x)=\sum_{k=0}^{\infty} p_k x^k$, by
\begin{equation}
	C_k[p(x), x] = p_k,
\end{equation}
Eq.~(\ref{condition3}) may be separated into multiple equations,
\begin{equation}
	C_{k}[\M[\H_{\eff}, t; t_0], 1/\omega] \stackrel{!}{=} C_{k}[\M[\H_{\rot}, t; t_0], 1/\omega] \qquad \forall k \in \mathbb{N}_0.
	\label{requirement}
\end{equation} % when determining $\M[\H_{\eff}]$ and $\M[\H_{\rot}]$
To ensure consistent counting of dimensions in $1/\omega$ before selecting the coefficients $C_k$, we follow the conventions for replacing parameters with time-like dimension used in the example calculations (\ref{mt0_example}) and (\ref{mt1_example}).  To be specific, we replace all instances of $t_c$ and $t_0$ following the rules $t_c \rightarrow \pi/\omega$ and $t_0 \rightarrow \beta_0/(2\omega)$ [cf.~Eqs.~(\ref{tc}) and (\ref{beta0})].  Furthermore, we replace those variables $t$ that have an impact on the dimensionality of the Magnus-Taylor expansion following the rule $t\rightarrow \beta/(2\omega)$ [cf.~Eq.~(\ref{beta})].  The arguments of unspecified functions such as $H_1(t)$ or $\dot H_1(t)$, which have no impact on the $1/\omega$-dimensionality of the coefficients in Eq.~(\ref{requirement}), are exempt from this replacement rule.  

To find the desired recurrence relation, first introduce a Hamiltonian decomposition,
\begin{equation}
	\H_{\eff}^{(N)}(t; t_0) = \sum_{k=0}^{N} \frac{h_k(t; t_0)}{\omega^k},
	\label{Hseries}
\end{equation}
similar to that given in the effective Hamiltonian series (\ref{Heff1}) except that here the summation ends at finite order $N$.  Our derivation of the time-dependent operators $h_k(t; t_0)$ can be viewed as constructing the effective Hamiltonians $\H_{\eff}^{(N)}$ ``from the bottom up,'' that is, we start with $N=0$ and then inductively increment $N \rightarrow N+1$ via a recursive procedure that determines $h_{N+1}$ as a function of $\H_{\eff}^{(N)}$. 

We begin our derivation by determining the zeroth-order effective Hamiltonian, $\H_{\eff}^{(0)}=h_0$.  To do this, we first evaluate the Magnus-Taylor expansion on the left-hand side (LHS) of Eq.~(\ref{requirement}) for the lowest order of $k=0$.  Using the notation introduced in Sec.~\ref{MT}, let us express the corresponding effective Hamiltonian (\ref{Hseries}) as $\H^{(N=0)}_{\eff}(t) = \H^{(0)}_{\eff}({\bf X}(t)) = h_0({\bf X}(t))$.  This (i) highlights its possible dependence on a set of parameters ${\bf X}(t)$, and (ii) conforms with our premise (\ref{HEffDD}) stating that the effective Hamiltonian does not explicitly depend on time.  Furthermore, from our discussion at the end of Sec.~\ref{MT} [below Eq.~(\ref{coefficients})] we take that for this order of $k=0$ we can truncate both the Magnus-Taylor series as given in Eq.~(\ref{MagnusTaylor}) and the associated Taylor series (\ref{Taylorexpansion}) at lowest order.  We thus find
\begin{eqnarray}
	\M[\H^{(0)}_{\eff}, t; t_0]  &=& \m_0[h_0, t; t_0] + \O(1/\omega) \nonumber \\
			&\stackrel{(\ref{mt0})}{=}& \frac1{t_c} \int_{t_0}^{t_0+t_c} \text d \tau h_0(\T[{\bf X}(\tau), t]) + \O(1/\omega) \nonumber \\
			&=& \frac1{t_c} \int_{t_0}^{t_0+t_c} \text d \tau h_0(t; t_0) + \O(1/\omega) \label{LHS_MTE-Taylor} \\
			&=& h_0(t; t_0) + \O(1/\omega).
			\label{M_of_Heff0}
\end{eqnarray}
In Eq.~(\ref{LHS_MTE-Taylor}) the integral has turned trivial, because the integrand has lost all dependence on the integration variable $\tau$.  Since this is a crucial step, we reiterate that this loss of $\tau$-dependence of the integrand comes about since the effective Hamiltonian's time dependence comes solely through the parameters ${\bf X}(\tau)$, whose Taylor expansion (\ref{Taylorexpansion}) has been truncated at lowest order, i.e., ${\bf X}(\tau) \simeq {\bf X}(t)$.  Equation (\ref{M_of_Heff0}) then implies that the zeroth-order coefficient
\begin{equation}
	C_{0}[\M[\H^{(0)}_{\eff}, t; t_0], 1/\omega] = h_0(t; t_0),
		\label{LHS0}
\end{equation}
indeed yields the lowest-order Hamiltonian term we seek at this step of the derivation.  

Now recall that in Eq.~(\ref{mt0_example}) we found that the lowest-order Magnus-Taylor expansion for the special-case Hamiltonian (\ref{Hrot2}) is equal to the RWA Hamiltonian.  This calculation can be generalized straightforwardly to yield the same result for the generic Hamiltonian (\ref{Hrot0}), which implies that the RHS of Eq.~(\ref{requirement}) is
\begin{equation}
	C_0[\M[\H_{\rot}, t; t_0], 1/\omega] = \H_{\RWA}(t).
	\label{RHS0}
\end{equation}
Combining Eq.~(\ref{requirement}) for $k=0$ with (\ref{LHS0}) and (\ref{RHS0}) we conclude
\begin{equation}
	\H_{\eff}^{(0)}(t; t_0) = h_0(t; t_0) = \H_{\RWA}(t).
	\label{h0}
\end{equation}
We have thus identified the lowest-order effective Hamiltonian with the RWA Hamiltonian (\ref{HRWAgeneric}).  

Next, let us discuss the recursion step $N \rightarrow N+1$.  First note that because the effective Hamiltonian has units of energy, or $\omega$ (since $\hbar=1$), each coefficient $h_k$ has units of $\omega^{k+1}$.  We now use this fact to relate the Magnus-Taylor expansions of two successive effective Hamiltonians $\H^{(N+1)}_{\eff}$ and $\H^{(N)}_{\eff}$ to one another.  Starting with the series (\ref{MagnusTaylor}) and separating out the $\m_0$ term,
\begin{eqnarray}
	\M[\H^{(N+1)}_{\eff}, t; t_0] &\stackrel{(\ref{Hseries})}{=}& 
			\m_0[\H_{\eff}^{(N)} + h_{N+1}/\omega^{N+1}, t; t_0] + \sum_{k=1}^{\infty} \m_k[\H_{\eff}^{(N)} + h_{N+1}/\omega^{N+1}, t; t_0] \nonumber \\ 
		&=& \m_0[\H_{\eff}^{(N)}, t; t_0] + \m_0[h_{N+1}/\omega^{N+1}, t; t_0] + \sum_{k=1}^{\infty} \m_k[\H_{\eff}^{(N)}, t; t_0] + \O(1/\omega^{N+2}) \nonumber\\  
		&=& \M[\H^{(N)}_{\eff}, t; t_0] + \m_0[h_{N+1}/\omega^{N+1}, t; t_0] + \O(1/\omega^{N+2}).
	\label{MagnusSplit}
\end{eqnarray}
The step from the first to the second line is nontrivial, and can be explained as follows.  First, the zeroth-order term $\m_0[\H, t; t_0]$, which is given in Eq.~(\ref{mt0}), is linear in its first argument, resulting in the sum of the terms $\m_0[\H^{(N)}_{\eff}, t; t_0]$ and $\m_0[h_{N+1}/\omega^{N+1}, t; t_0]$.  Second, when ignoring $h_{N+1}/\omega^{N+1}$ inside $\m_{k}[\H^{(N+1)}_\eff, t; t_0]$ with $k \geq 1$ the only terms we do not keep explicitly are due to the commutator of $\H^{(N)}_{\eff}$ and $h_{N+1}$, or those due to the commutator of $h_{N+1}$ with itself (at different times).  Consider, for example, the commutator of $h_0$ and $h_{N+1}$, which is of lowest order of the terms neglected in this step.  Recalling that, as noted directly above Eq.~(\ref{MagnusSplit}), $h_k$ has units of $\omega^{k+1}$, this commutator has units of $\omega^{N+3}$, and since the Magnus-Taylor expansion itself has units of $\omega$, this lowest-order correction term must be proportional to $1/\omega^{N+2}$.  Finally, all neglected terms are collected in $\O(1/\omega^{N+2})$ since the one just discussed is that of lowest order.  

We now assume that for a given $N\in\mathbb{N}_0$ we have found the effective Hamiltonian $\H_{\eff}^{(N)}$, which satisfies the requirements (\ref{requirement}) for all $k \leq N$.  It turns out that the only nontrivial requirement (\ref{requirement}) for the next order is that for $k=N+1$, because the requirements for all $k \leq N$ are automatically fulfilled by $\H_{\eff}^{(N+1)}$.  To see why this is the case, we use the relation (\ref{MagnusSplit}) established above to simplify the LHS of Eq.~(\ref{requirement}) for the Hamiltonian $\H_{\eff}^{(N+1)}$ and $k\leq N$,
\begin{eqnarray}
	C_{k}[\M[\H_{\eff}^{(N+1)}, t; t_0], 1/\omega] &\stackrel{(\ref{MagnusSplit})}{=}& C_{k}[\M[\H^{(N)}_{\eff}, t; t_0] + \m_0[h_{N+1}, t; t_0]/\omega^{N+1}, 1/\omega] \nonumber \\
				&=& C_{k}[\M[\H_{\eff}^{(N)}, t; t_0], 1/\omega].
\end{eqnarray}
This result implies that the LHS of Eq.~(\ref{requirement}) for the Hamiltonian $\H_{\eff}^{(N+1)}$ and $k \leq N$ can be reduced to that of $\H_{\eff}^{(N)}$, which at this point, as noted above, is assumed to fulfill the requirement (\ref{requirement}) for all $k \leq N$.  

Moving on to the next order coefficient $C_k$ with $k=N+1$, we again use Eq.~(\ref{MagnusSplit}) to simplify the LHS of the requirement (\ref{requirement}),
\begin{eqnarray}
	C_{N+1}[\M[\H_{\eff}^{(N+1)}, t; t_0], 1/\omega] &\stackrel{(\ref{MagnusSplit})}{=}& C_{N+1}[\M[\H^{(N)}_{\eff}, t; t_0] + \m_0[h_{N+1}/\omega^{N+1}, t; t_0], 1/\omega] \nonumber \\
				&=& C_{N+1}[\M[\H^{(N)}_{\eff}, t; t_0], 1/\omega] + C_{N+1}[\m_0[h_{N+1}/\omega^{N+1}, t; t_0], 1/\omega] \nonumber \\
				&=& C_{N+1}[\M[\H^{(N)}_{\eff}, t; t_0], 1/\omega] + h_{N+1}(t; t_0). \label{N+1}
\end{eqnarray}
When going from the first to the second line, we have used the linearity of the coefficient operator $C_k$.  In the step leading to the third line, we have used the fact that the Taylor series (\ref{Taylorexpansion}) is to be truncated at zeroth order since any higher-order terms are at an increased order in $1/\omega$.  Similar to the evaluation in Eq.~(\ref{M_of_Heff0}), this truncation has the effect of $h_{N+1}$ losing all dependence on the integration variable, thus rendering the evaluation of the Magnus-Taylor term $\m_0$ trivial.  

Finally, combining the result of Eq.~(\ref{N+1}) with the requirement (\ref{requirement}) for $k=N+1$, and solving for the new coefficient $h_{N+1}$ we obtain our central recurrence relation
\begin{align}
	h_{N+1}(t; t_0) \ = \ & C_{N+1}[\M[\H_{\rot}, t; t_0] - \M[\H_{\eff}^{(N)}, t; t_0], 1/\omega] \nonumber\\
	\stackrel{(\ref{Hseries})}{\Longrightarrow}	\quad \Aboxed{ h_{N+1}(t; t_0) \ = \ & C_{N+1}\left[ \M[\H_{\rot}, t; t_0] - \M\left[\sum_{k=0}^{N} \frac{h_k(t; t_0)}{\omega^k}, t; t_0\right], 1/\omega \right] }.
	\label{recursion}
\end{align}
The effective Hamiltonian (\ref{Hseries}) is defined by this equation together with the starting point, $h_0 = \H_{\RWA}$, as given in Eq.~(\ref{h0}).  

The recursive procedure for calculating the effective Hamiltonian $\H_{\eff}^{(N)}$ for some order $N$ can thus be summarized as follows.  Beginning from the lowest-order Hamiltonian (\ref{h0}), $\H_{\eff}^{(0)}(t; t_0) = \H_{\RWA}(t)$, all higher terms $h_k$ with $k \geq 1$ are then obtained via repeated evaluation of the recurrence relation (\ref{recursion}).  We reiterate that, as stated below Eq.~(\ref{requirement}), when deploying this procedure it is essential that after computing each Magnus-Taylor expansion in Eq.~(\ref{recursion}) and before taking the coefficient $C_{N+1}$, all temporal parameters are replaced by their equivalent dimensionless parameters, unless they are arguments of the unknown envelope $H_1(t)$ or its derivatives [see also our example calculation of the Magnus-Taylor expansion given above in Eqs.~(\ref{mt1_example})-(\ref{MagnusTaylor1})].  In the following section we use this procedure to calculate a first-order effective Hamiltonian.

\subsection{Example Calculation:  Effective Hamiltonian of Order 1/$\omega$}
\label{HeffExample}

Let us now use our recursion procedure to calculate the effective Hamiltonian up to first order in $1/\omega$, $\H_{\eff}^{(1)}$.  As before in the example calculation of the Magnus-Taylor expansion in Sec.~\ref{MT}, we once again concentrate on the simple rotating-frame Hamiltonian (\ref{Hrot2}), $\mathcal H_{\rot}(t) = (H_1(t)/4) ( \sigma_x + \cos(2\omega t) \sigma_x - \sin(2\omega t)\sigma_y)$.  

We start the construction of this effective Hamiltonian (\ref{Hseries}) for $N=1$ with the lowest order which, as stated in Eq.~(\ref{h0}), is the Hamiltonian in the rotating wave approximation (RWA),
\begin{equation}
	\H_{\eff}^{(N=0)}(t; t_0) = h_0(t; t_0) = \H_{\RWA}(t) \stackrel{(\ref{HRWA})}{=} \frac{H_1(t)}4 \sigma_x.
	\label{Heff(0)}
\end{equation}
For the next order,
\begin{equation}
	\H_{\eff}^{(N=1)}(t; t_0) \stackrel{(\ref{Hseries})}{=} h_0(t; t_0) + h_1(t; t_0)/\omega,
	\label{Heff(1)}
\end{equation}
we determine the Hamiltonian coefficient $h_1$ using the recurrence relation (\ref{recursion}) for $N=0$,
\begin{eqnarray}
	h_{1}(t; t_0) &=& C_{1}[\M[\H_{\rot}, t; t_0] - \M[\H_{\RWA}(t), t; t_0], 1/\omega],
		\label{recursion0}
\end{eqnarray}
where we have already used $\H_{\eff}^{(N=0)} = \H_{\RWA}$.  Combining the first-order Magnus-Taylor expansion of $\H_{\rot}$, calculated in Eqs.~(\ref{mt0_example}) and (\ref{mt1_example}), results in
\begin{equation}
	C_{1}[\M[\H_{\rot}, t; t_0], 1/\omega] = \frac{H_1(t)^2}{32} (1-2\cos\beta_0) \sigma_z + \frac{\dot H_1(t)}{8} ((\pi + \beta_0 - \beta + \sin \beta_0) \sigma_x + \cos \beta_0 \sigma_y). % \nonumber \\
\end{equation}

The Magnus-Taylor expansion of $\H_{\RWA}$ up to order $1/\omega$, similar to the exemplary calculation of the Magnus-Taylor expansion up to the same order of the Hamiltonian $\H_{\rot}$ in Eq.~(\ref{MagnusTaylor1}), consists of only the first two terms of the series (\ref{MagnusTaylor}),
\begin{equation}
	\M[\H_{\RWA}(t), t; t_0] = \m_0[\H_{\RWA}(t), t; t_0] + \m_1[\H_{\RWA}(t), t; t_0] + \O(1/\omega^2).
\end{equation}
As discussed at the end of Sec.~\ref{MT}, for the first-order Magnus-Taylor expansion the Taylor series (\ref{Taylorexpansion}) of $X_i = H_1$ can be truncated at $k=1$.  Using the definition of $\m_0$ as given by Eq.~(\ref{mt0}), we thus obtain
\begin{eqnarray}
	\m_0[\H_{\RWA}(t), t; t_0] &=& \frac 1{t_c} \int_{t_0}^{t_0+t_c} \text d \tau \left(\frac{H_1(t)}4 + \frac{\dot H_1(t)}4 (\tau-t) + \O((\tau-t)^2)\right) \sigma_x \nonumber \\
			&=& \frac{H_1(t)}4 \sigma_x + \frac{\dot H_1(t)}{8 \omega} (\pi + \beta_0 - \beta)\sigma_x + \O(1/\omega^2).
\end{eqnarray}
The next term $\m_1$, given by Eq.~(\ref{mt1}), vanishes since the zeroth-order effective Hamiltonian (\ref{Heff(0)}) is proportional to $\sigma_x$ at all times, and therefore commutes with itself at all times,
\begin{equation}
	\m_1[\H_{\RWA}(t), t; t_0] = 0.
	\label{mt1_2nd-example}
\end{equation}

Combining Eqs.~(\ref{recursion0}) through (\ref{mt1_2nd-example}), we find
\begin{equation}
	h_{1}(t; t_0) 
			= \frac{H_1(t)^2}{32} (1-2\cos\beta_0) \sigma_z + \frac{\dot H_1(t)}{8} (\sin \beta_0 \sigma_x + \cos \beta_0 \sigma_y).
\end{equation}
Together with $h_0$ given in Eq.~(\ref{Heff(0)}), we conclude that the first-order effective Hamiltonian (\ref{Heff(1)}) is given by
\begin{equation}
	\H_{\eff}^{(1)}(t; t_0) = \frac{H_1(t)}4 \sigma_x + \frac{H_1(t)^2}{32\omega} (1-2\cos\beta_0) \sigma_z + \frac{\dot H_1(t)}{8 \omega} (\sin \beta_0 \sigma_x + \cos \beta_0 \sigma_y).
\end{equation}
As noted in the Introduction, generic effective Hamiltonians up to second order in $1/\omega$ can be found in \ref{appendix:examples}.

\subsection{Simplified Computation Method for Time-Independent Envelope}
\label{Heffconst}

For the case of a constant envelope $H_1(t) \equiv H_1$, the above procedure for calculating the effective Hamiltonian turns relatively simple.  In fact, we now show that for this special case the effective Hamiltonian can be computed via a regular Magnus expansion of the rotating-frame Hamiltonian (\ref{Hrot0}),
\begin{equation}
	\qquad \qquad \qquad \qquad \H_{\eff}(t_0) = \overline \H_{\rot},\qquad \qquad \qquad \qquad (H_1(t) = H_1).
	\label{HEffClaim}
\end{equation}
Recall that in the notation for the Magnus expansion, introduced in Sec.~\ref{M} [see Eqs.~(\ref{UM1})-(\ref{Hb2})], the dependence on $t_0$ and $n$ is suppressed.  The Magnus expansion in Eq.~(\ref{HEffClaim}) is to be  taken on the fundamental Magnus interval $[t_0, t_0 + t_c)$, so that the first three terms of the Magnus series (\ref{MagnusSeries}) are given by Eqs.~(\ref{Hb0})-(\ref{Hb2}) for the case of $n=1$.  

It is not difficult to derive Eq.~(\ref{HEffClaim}) given our central recurrence formula (\ref{recursion}) derived in Sec.~\ref{procedure}.  This is because for a constant envelope several simplifications arise.  First recall that for this case, as discussed in Sec.~\ref{intro:Time Independent Drives}, the effective Hamiltonian introduced as the series representation (\ref{Heff0}) itself is time independent.  Terminating this series at order $N$ similar to the generic effective Hamiltonian series (\ref{Hseries}), we have
\begin{equation}
	\H_{\eff}^{(N)}(t_0) = \sum_{k=0}^{N} h_k(t_0) (1/\omega)^k.
	\label{HseriesConstant}
\end{equation}
Second, the Magnus-Taylor expansion (\ref{MagnusTaylor}) reduces to a regular Magnus expansion (\ref{MagnusSeries}) for the Magnus interval $[t_0, t_0 + t_c)$, because the Taylor series (\ref{Taylorexpansion}) for the only occurring parameter $X_1(\tau) = H_1(\tau) = H_1$ terminates at lowest order.  As a consequence, the recurrence relation (\ref{recursion}) for the effective Hamiltonian simplifies to
\begin{equation}
	h_{N+1}(t_0) = C_{N+1}\left[\overline \H_{\rot} - \overline \H_{\eff}^{(N)}, 1/\omega \right],
	\label{recursionSimple1}
\end{equation}
where we have replaced both Magnus-Taylor expansions in Eq.~(\ref{recursion}) by regular Magnus expansions, and further replaced $h_{N+1}(t; t_0)$ with $h_{N+1}(t_0)$.  

Now note that since the effective Hamiltonian $\H_\eff(t_0)$ is independent of time $t$, it commutes with itself at arbitrary times.  For this reason all terms but that of lowest order in the Magnus series (\ref{MagnusSeries}) vanish, i.e., $\overline \H_\eff = \sum_{k=0}^\infty \overline \H_\eff^{(k)} \equiv \overline \H_\eff^{(0)}$.  The lack of dependence on time $t$ of $\H_{\eff}$ further implies that this lowest-order term, which is a straightforward average over the fundamental Magnus interval as given in Eq.~(\ref{Hb0}) for $n=1$, is simplified trivially as $\overline \H_\eff^{(0)} = \frac1{t_c} \int_{t_0}^{t_0 + t_c} \text d \tau \H_{\eff}(t_0) \equiv \H_{\eff}(t_0)$.  Since the same simplification applies to the Magnus expansion of the (time-independent) effective Hamiltonian series (\ref{HseriesConstant}), the $1/\omega^{N+1}$ coefficient of the effective Hamiltonian appearing in Eq.~(\ref{recursionSimple1}) trivially evaluates to zero,
\begin{equation}
	C_{N+1}\left[\overline \H_{\eff}^{(N)}, 1/\omega \right] \equiv C_{N+1}\left[\H_{\eff}^{(N)}, 1/\omega \right] = 0.
	\label{CMHEffConstant}
\end{equation}
Combining Eqs.~(\ref{recursionSimple1}) and (\ref{CMHEffConstant}) we find $h_{N+1}(t_0) = C_{N+1}\left[\overline \H_{\rot}, 1/\omega \right]$, which, when combined with Eq.~(\ref{HseriesConstant}), results in the simplified computation method (\ref{HEffClaim}) of the effective Hamiltonian.  

While the computation of the effective Hamiltonian for constant drive amplitudes is comparatively simple, we emphasize that computing the effective time evolution for a freely chosen gauge parameter $\beta_0$ [corresponding to $t_0 = \beta_0/(2\omega)$ for the time evolution operator $U_{t_0}$ given in Eq.~(\ref{Ueff})] requires the use of kick operators, as discussed in Sec.~\ref{intro:gauge_kicks}.  The derivation of these kick operators is presented in the next section.

\section{Kick Operators for non-Smooth Drives}
\label{nonanalytic}

The procedure for determining our effective Hamiltonians developed in Sec.~\ref{effectiveH} is based on the premise that the drive envelope $H_1(t)$ is completely smooth, i.e., all its derivatives exist.  This property is required to satisfy the assumption that the envelope changes sufficiently slowly in time as expressed by Eq.~(\ref{assumption}).  However, as noted in Sec.~\ref{intro:gauge_kicks}, realistic drive envelopes often do not fulfill this requirement.  

If the $k$th derivative of the envelope, $\dersub{H_1}{k}$, diverges, our effective Hamiltonian derivation cannot be carried through straightforwardly beyond order $1/\omega^{k-1}$.  To account for more generic drive envelopes satisfying the generalized assumption (\ref{assumption2}), which allows for divergences in the form of a $\delta$-function, we now introduce kick operators.  To compute the correct time evolution, we make use of a kick operator for each time $t_d$ at which $\dersub{H_1}{k}$ diverges for some $k$.  

\begin{figure}[tb]
	\includegraphics[width=\textwidth]{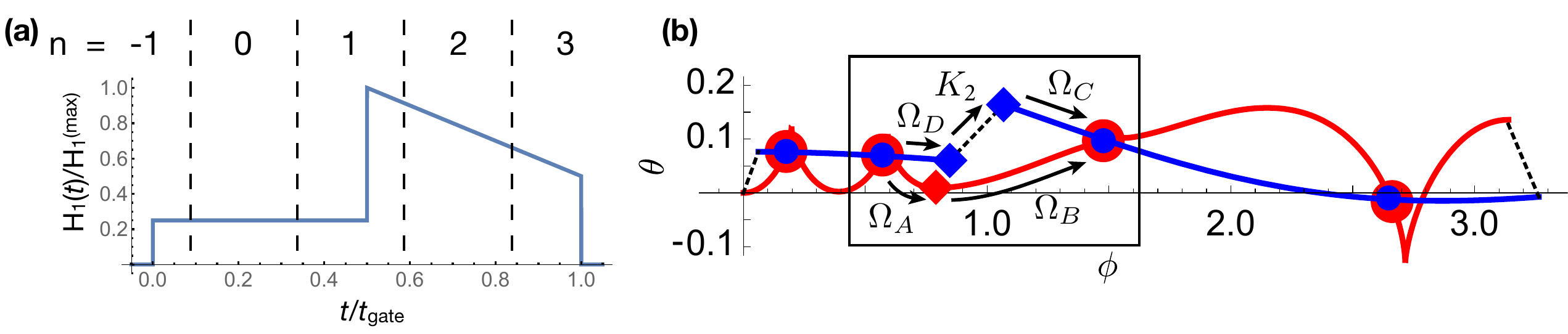}
	\caption{Qubit trajectories (as usual in the rotating frame) for piecewise-analytic driving envelope.  (a) Envelope defined in Note \cite{H_discon}, whose derivative diverges at times $t_d=0$ with $n=-1$ [cf.~Eq.~(\ref{nthInterval})], $t_d=t_{\gate}/2$ with $n=1$ and $t_d=t_{\gate}$ with $n=3$.  (b)  The axes of the plot represent the latitude $\theta$ and longitude $\phi$ of a rotated Bloch sphere where $|+\rangle = (|0\rangle + |1\rangle)/\sqrt{2}$ defines the north pole $(\phi, \theta) = (0, 0)$ (see also Note \cite{rotatedBS}), at which the qubit is initialized.  Shown trajectories are exact (red) and effective (blue) for the gauge parameter $\beta_0 = 0.45\times2\pi$.  Instantaneous displacements, caused by kick operators, are shown as dashed lines that connect the effective trajectories for $t<t_d$ and $t>t_d$ [cf.~Eq.~(\ref{U_with_kicks})].  For the second displacement, labeled $K_{2}$, the endpoints of each trajectory before and after $t_d = t_{\gate}/2$ are indicated by diamonds.  Surrounding generators of exact and effective time evolution operators across intervals (\ref{anterior}) and (\ref{posterior}) are labeled $\Omega_A$ through $\Omega_D$.}
	\label{fig:jumpop-trace}
\end{figure}

Figure \ref{fig:jumpop-trace} exemplifies the role played by kick operators for a non-smooth envelope function.  For the case of the envelope $H_1(t)$ shown in Fig.~\ref{fig:jumpop-trace}(a), the first derivative $\dot H_1(t)$ diverges at the three times $t_d = 0$, $t_d=t_{\gate}/2$, and $t_d = t_{\gate}$.  Figure \ref{fig:jumpop-trace}(b) shows both the exact (red) and effective (blue, including dashed lines) trajectories similar to those shown in Figs.~\ref{trajectories} and \ref{gauge_and_kicks}, but which are plotted in a $(\phi, \theta)$-polar representation of the Bloch sphere surface for a qubit initialized at the north pole, $(\phi, \theta)=(0,0)$.  There are three instantaneous displacements in the effective trajectory shown in the figure (indicated by dashed lines), and each of these is the result of applying a kick operator at the corresponding time of divergence of the envelope's derivative.  The second time of divergence, $t_d=t_{\gate}/2$, serves as an illustrative example in the following discussion.  

For our derivation of the kick operator we assume, in accordance with the extended assumption (\ref{assumption2}), that the envelope $H_1(t)$ diverges at no more than one time $t_d$ within a Magnus interval,
\begin{equation}
	[t_0+nt_c, t_0+(n+1)t_c),
	\label{nthInterval}
\end{equation}
which is defined by a time offset $t_0\in[0, t_c)$ and an integer $n$.  This is a reasonable assumption whenever the time $t_c=\pi/\omega$ is short compared to the gate duration $t_\gate$.  Now consider a generic envelope $H_1(t)$ exhibiting a discontinuity at time $t_d$ in an interval (\ref{nthInterval}).  We begin our derivation by introducing Hamiltonians $\H^{<}_{\rot}$ and $\H^{>}_{\rot}$ similar to Eq.~(\ref{Hrot0}) with envelopes $H_1(t) = H^<_{1}(t)$ and $H^>_{1}(t)$, respectively, all of which we define below in a way that they are smooth on the \emph{entire} interval (\ref{nthInterval}).  In the anterior part of the interval (\ref{nthInterval}),
\begin{equation}
	[t_0+nt_c, t_d),
	\label{anterior}
\end{equation}
the envelope $H^<_{1}(t)$ is given by $H_1(t)$, while in the posterior part,
\begin{equation}
	[t_d, t_0+(n+1)t_c),
	\label{posterior}
\end{equation}
the same envelope $H^<_{1}(t)$ is equal to its analytic continuation in that region.  Conversely, the envelope $H^>_{1}(t)$ equals $H_1(t)$ in the posterior part of the interval (\ref{nthInterval}), while in the anterior part it is equal to its analytic continuation.  With reference to the regular rotating-frame Hamiltonian (\ref{Hrot0}), we introduce Hamiltonians $\mathcal H^{\lessgtr}_{\rot}$ that are smooth on the entire Magnus interval (\ref{nthInterval}),
\begin{equation}
	\mathcal H^{\lessgtr}_{\rot}(t) = \frac{H^{\lessgtr}_{1}(t)}4 ( \cos(\phi)\sigma_x + \cos(2\omega t + \phi) \sigma_x + \sin(\phi)\sigma_y - \sin(2\omega t + \phi)\sigma_y) + \frac\Delta2 \sigma_z.
	\label{piecewiseH}
\end{equation}
With reference to the effective Hamiltonian (\ref{Heff1}), we similarly introduce effective Hamiltonians,
\begin{equation}
	\H^{\lessgtr}_{\eff}(t; t_0) = \sum_{k=0}^{\infty} \frac{h^{\lessgtr}_{k}(t; t_0)}{\omega^k},
	\label{piecewiseHeff}
\end{equation}
each of which can then be obtained via the method of Sec.~\ref{effectiveH} using the respective smooth Hamiltonian $\H^{\lessgtr}_{\rot}$.  Therefore, both effective Hamiltonians $\H^{\lessgtr}_{\eff}$ are also smooth on the entire Magnus interval (\ref{nthInterval}).  

The appearance of a kick operator $K_j(t_d; t_0)$ at the $j$th time of divergence $t_d$ is a direct consequence of the condition that the effective and exact trajectories are to coincide at the boundaries of the Magnus interval (\ref{nthInterval}).  In fact, we use this condition to derive the $j$th kick operator via its corresponding unitary operator $e^{K_j}$, which provides an impulse connecting the effective trajectories for times $t<t_d$ and $t>t_d$.  We define generators $\Omega_i$ with $i = A, B, C, D$ on a Magnus interval (\ref{nthInterval}) via the framed section shown in Fig.~\ref{fig:jumpop-trace}(b), for which $n=1$ [cf.~Fig.~\ref{fig:jumpop-trace}(a)], $j=2$ and $t_d=t_{\gate}/2$.  That is, $e^{\Omega_A}$ and $e^{\Omega_B}$ correspond to the \emph{exact} time evolutions across the anterior interval (\ref{anterior}) and the posterior interval (\ref{posterior}), respectively,  while $e^{\Omega_D}$ and $e^{\Omega_C}$ correspond to the respective \emph{effective} time evolutions.  Taking into account the action of the kick operator $K_j$ for the effective time evolution, the demand on a stroboscopic time evolution [cf.~the equality of the effective and exact time evolutions as stated in Eq.~(\ref{goal0})] implies equal propagators across the interval (\ref{nthInterval}), i.e.,
\begin{equation}
	U_{t_0}(t_0+(n+1)t_c, t_0+nt_c) =  e^{\Omega_C} e^{K_j(t_d; t_0)} e^{\Omega_D} \stackrel{!}{=} e^{\Omega_B} e^{\Omega_A}. \label{construction}
\end{equation}
Solving for the kick operator,
\begin{equation}
	e^{K_j(t_d; t_0)} = e^{-\Omega_C} e^{\Omega_B} e^{\Omega_A} e^{-\Omega_D}.
	\label{eq:omega_j-defintion}
\end{equation}
we are then able to compute $K_j$ perturbatively.  To do this, we first evaluate $\Omega_A$ through $\Omega_D$ as a series expansion in $1/\omega$ using the Magnus-Taylor expansion (introduced in Sec.~\ref{MT}) of the Hamiltonians (\ref{piecewiseH}) and (\ref{piecewiseHeff}), and then apply the Baker-Campbell-Hausdorff formula to Eq.~(\ref{eq:omega_j-defintion}).  

Recall that we defined the Magnus-Taylor expansion $\M[\H, t; t_0]$ in Eq.~(\ref{MagnusTaylor}) for (i) a Hamiltonian $\H$, (ii) a reference time $t$ for the associated Taylor series (\ref{Taylorexpansion}) and (iii) a time offset $t_0$, which determines the fundamental Magnus interval (\ref{fundamental}), $[t_0, t_0+t_c)$.  To compute the Magnus-Taylor expansion for a generic interval $[t_a, t_b)$ we now introduce an extended notation,
\begin{equation}
	\M[\H, t; t_a, t_b] = \sum_{k = 0}^{\infty} \m_k[\H, t; t_a, t_b],
	\label{MagnusExtended}
\end{equation}
in which both interval boundaries $t_a$ and $t_b$ are given explicitly.  The lowest two terms in the summation of Eq.~(\ref{MagnusExtended}), $m_0$ and $m_1$, are respectively given by Eqs.~(\ref{mt0}) and (\ref{mt1}) upon replacing the integral bounds via $t_0\rightarrow t_a$ and $t_0+t_c \rightarrow t_b$.  Below we replace the envelope $H_1(t)$ and its derivatives by Taylor series with respect to the time of the divergence; that is, we use the Taylor series (\ref{Taylorexpansion}) with $t = t_d$ to determine Magnus-Taylor expansions of the form $\M[\H, t_d; t_a, t_b]$.  

From the definition of the propagators $e^{\Omega_A}$ through $e^{\Omega_D}$ [cf.~Fig.~\ref{fig:jumpop-trace}(b)], it is clear that the required Magnus-Taylor expansions are taken for the anterior interval (\ref{anterior}) or the posterior interval (\ref{posterior}).  Furthermore, similar to the time evolution (\ref{UMT}) across an entire Magnus interval expressed via the Magnus-Taylor expansion (\ref{MagnusTaylor}), the operators $\Omega_A$ through $\Omega_D$ can be expressed via the Magnus Taylor expansion with extended notation given in Eq.~(\ref{MagnusExtended}).  Because of this we have
\begin{eqnarray}
	\Omega_A &=& -i \M[\H_{\rot}^<, t_d; t_0 + n t_c, t_d]\ (t_d - (t_0 + n t_c)),
			\label{OmegaA}\\
	\Omega_B &=& -i \M[\H^>_{\rot}, t_d; t_d, t_0+(n+1)t_c]\ (t_0 + (n+1) t_c - t_d),\\
	\Omega_C &=& -i \M[\H^>_{\eff}, t_d; t_d, t_0+(n+1)t_c]\ (t_0 + (n+1) t_c - t_d), \\
	\Omega_D &=& -i \M[\H^<_{\eff}, t_d; t_0+nt_c, t_d]\ (t_d - (t_0 + n t_c)),
	\label{OmegaD}
\end{eqnarray}
where again $n=1$ for the time evolution within the framed section shown in Fig.~\ref{fig:jumpop-trace}.  Below we express the kick operators using a dimensionless parameter that replaces the time of the divergence $t_d$,
\begin{equation}
	\qquad \qquad \qquad \beta_d = 2\omega t_d,
	\label{betad}
\end{equation}
which is defined in analogy to the gauge parameter $\beta_0 = 2\omega t_0$.  

Similar to the effective Hamiltonian expansion (\ref{Heff1}), we now expand the kick operator $K_j(t_d; t_0)$ as a power series in $1/\omega$,
\begin{equation}
	K_j(t_d; t_0) = \sum_{k=1}^\infty \frac{K_j^{(k)}(t_d; t_0)}{\omega^k}.
	\label{K}
\end{equation}
The terms $K_j^{(k)}$ are obtained by combining Eq.~(\ref{eq:omega_j-defintion}) with Eqs.~(\ref{OmegaA})-(\ref{OmegaD}) and, as noted above, applying the Baker-Campbell-Hausdorff formula.  Similar to the example calculations in Sec.~\ref{MT}, in order to determine each term $K_j^{(k)}$ with its appropriate dimensionality in $1/\omega$, in this calculation we replace the time parameters $t_c$, $t_0$ and $t_d$ by their respective dimensionless parameters [see Eqs.~(\ref{tc}), (\ref{beta0}) and (\ref{betad})].  The first two terms of this expansion are then given by
\begin{eqnarray}
	K_j^{(1)}(t_d; t_0) & = & i \frac{H_1^<-H_1^>}{8}[(\sin(\beta_0+\phi)-\sin(\beta_d+\phi))\sigma_x+(\cos (\beta_0+\phi) -\cos(\beta_d+\phi))\sigma_y], \label{K1} \\
	K_j^{(2)}(t_d; t_0) & = & i\frac{(H_1^<)^2-(H_1^>)^2}{64}(\sin(\beta_d -\beta_0)-2 \sin(\beta_d +2 \phi)+2 \sin(\beta_0+2 \phi))\sigma_z \nonumber \\
			&& + i\frac{\Delta (H_1^<-H_1^>)}{16}[(\sin(\beta_d +\phi)-\sin(\beta_0+\phi))\sigma_x+(\cos(\beta_d +\phi)-\cos(\beta_0+\phi))\sigma_y] \nonumber \\
			&& + i\frac{\dot H_1^<-\dot H_1^>}{16}[(\cos(\beta_0+\phi)-\cos(\beta_d +\phi))\sigma_x+(\sin(\beta_d +\phi)-\sin(\beta_0+\phi))\sigma_y], \label{K2}
\end{eqnarray}
where $H^\lessgtr_{1}$ and $\dot{H}^\lessgtr_{1}$ are shorthand for the Taylor coefficients $H^\lessgtr_{1}(t_d)$ and $\dot{H}^\lessgtr_{1}(t_d)$, respectively.  The third order coefficient $K_j^{(3)}$, which is the lowest non-vanishing kick operator term for the envelope with sinusoidal ramp \cite{envelopeFunction}, is available on request.  

We note that an interesting situation occurs when the time of divergence coincides with the edge of a Magnus interval (\ref{nthInterval}), i.e., if $t_d = t_0 + n t_c$ for an integer $n$.  In such a case the above derivation of the kick operators turns trivial, starting with the construction of the anterior and posterior intervals and the corresponding envelope functions $H_1^{\lessgtr}$ and Hamiltonians $\H_{\rot}^\lessgtr$ and $\H_{\eff}^\lessgtr$ [cf.~Eqs.~(\ref{anterior})-(\ref{piecewiseHeff})].  As a consequence, the kick operator $K_j(t_d = t_0 + n t_c; t_0) = 0$ vanishes.  [This is reflected by the fact that $K_j^{(1)}$ and $K_j^{(2)}$, given above explicitly, vanish for this case, since $t_d = t_0 + n t_c$ implies $\beta_d = \beta_0 \pmod{2\pi}$.]  The total number of nontrivial kick operators may thus be minimized by an appropriate choice of the gauge parameter $\beta_0$.  

In this section we have thus extended our theory for restricted drives, whose slowly varying amplitude functions are completely smooth, to more generic amplitude functions that fulfill the generalized assumption (\ref{assumption2}).  The generator of the time evolution operator, given in Eq.~(\ref{Heff_generic}), is a sum of the smooth parts of the effective Hamiltonians, defined via piecewise smooth envelopes and obtained using the recursive procedure derived in Sec.~\ref{effectiveH}, and kick operators $i K_{j}(t_d; t_0) \delta(t-t_d)$ for the $j$th  discontinuity, where $K_j$ is given by Eq.~(\ref{K}).  For example, in the case of the envelope shown in Fig.~\ref{fig:jumpop-trace}(a) the combined effective Hamiltonian reads
\begin{eqnarray}
	\H_{\eff}(t; t_0) &=& \H_\eff^<(t_0) \Theta(t) \Theta(t_\gate/2-t) + \H_\eff^>(t; t_0) \Theta(t-t_\gate/2) \Theta(t_\gate-t) \nonumber \\
	&& + i K_1(0; t_0) \delta(t) + i K_{2}(t_\gate/2; t_0) \delta(t-t_\gate/2) + i K_{3}(t_\gate; t_0) \delta(t-t_\gate).
	\label{HNon-analytic}
\end{eqnarray}
Here we have taken into account the fact that the envelope for times $t\in[0, t_\gate/2]$ is constant, so that for this interval the effective Hamiltonian is independent of the current time $t$.  

The effective time evolution operator,
\begin{equation}
	U_{t_0}(t, 0^{-}) = \mathcal T e^{-i\int_{0^-}^t \d \tau \H_\eff(t; t_0)},
\end{equation}
[cf.~Eq.~(\ref{Ueff})] for the complete pulse with the envelope shown in Fig.~\ref{fig:jumpop-trace}(a) is given by
\begin{equation}
	U_{t_0}(t_{\gate}^+, 0^{-}) = e^{K_3(t_\gate; t_0)} \left(\mathcal{T} e^{-i \int_{t_\gate/2}^{t_\gate} \d\tau \H_{\eff}^>(\tau; t_0)} \right) e^{K_2(t_\gate/2; t_0)} e^{-i\H_\eff^<(t_0) t_\gate/2} e^{K_1(0; t_0)}.
	\label{U_with_kicks}
\end{equation}
Here the evolution is considered from initial time $t_i = 0^-$ slightly before the beginning of the pulse to final time $t_{\gate}^+$ slightly after the end of the pulse, in order to include all $\delta$-functions that appear in the Hamiltonian (\ref{HNon-analytic}).  Recall that, as introduced in Sec.~\ref{intro:gauge_kicks}, $\beta_0 = 2\omega t_0\in[0,2\pi)$ is called a gauge parameter because its choice does not alter the result of the time evolution operator $U_{t_0}(t_{\gate}^+, 0^{-})$ over the entire pulse.  

Recall that in Sec.~\ref{intro:gauge_kicks} we illustrated the action of kick operators on the basis of the square pulse whose envelope function is shown in Fig.~\ref{trajectories}(b).  The corresponding trajectories, exemplified on the left of Fig.~\ref{gauge_and_kicks} for several gauge parameters $\beta_0$, can be obtained from a combined effective Hamiltonian containing kick operators and the resulting effective time evolution operator [these operators are similar to those in Eqs.~(\ref{HNon-analytic}) and (\ref{U_with_kicks})].  This time evolution operator then includes instantaneous displacements at the beginning and end of each trajectory.  We note that the effective qubit trajectories would be the same if the constant envelope was ``always on," i.e., $H_1(t) \equiv H_1$ for all times $t$, assuming the same initial condition $|\psi(t=0)\rangle=|0\rangle$ as that used in Fig.~\ref{gauge_and_kicks}.  For such a pulse, the kick operators derived in this section can similarly be used to determine the displacement that connects one set of stroboscopic state vectors $|\psi(t_n)\rangle$ with $t_n\in \{t_0, t_0+t_c, \ldots \}$ to another set that contains state vectors $|\psi(t_n')\rangle$ with $t_n'\in \{t_0', t_0'+t_c, \ldots \}$.

\section{Conclusions}
\label{conclusions}

The objective of the present work has been to study the time evolution of a linearly driven qubit in the regime of strong driving, i.e., for field strengths smaller than or comparable to the drive frequency $\omega$, or $|H_1(t)|\lesssim \omega$, with special emphasis on the consequences of the envelope being time dependent.  For strong and time-dependent driving, the errors of the predicted time evolution using standard methods---such as the rotating wave approximation combined with Bloch-Siegert shifts---are appreciable.  We have addressed this problem by introducing an \emph{effective Hamiltonian} which generates a stroboscopic time evolution; that is, the time evolution operator due to our effective Hamiltonian agrees with the exact time evolution operator at points equally spaced in time.  The time difference between two points of agreement is equal to the duration of the Bloch-Siegert oscillations, or the drive period in the rotating frame.  Since our effective Hamiltonian generalizes the rotating wave approximation and allows one to approximate the exact trajectory to arbitrary accuracy, we call our theory the exact rotating wave approximation.  

The effective Hamiltonian has been obtained as a power series in the inverse drive frequency.  In order to compute the coefficients of this series in a systematic fashion, we have introduced the \emph{Magnus-Taylor expansion}, a new method for performing time-dependent perturbation theory that utilizes both a Magnus expansion and a Taylor series.  This Magnus-Taylor expansion has allowed us to derive a recurrence relation that determines the set of operator coefficients that make up the effective Hamiltonian.  Assuming an envelope that varies only slightly on the time scale of the drive period [cf.~Eq.~\ref{assumption2}], our effective Hamiltonian---as in the case of the Hamiltonian in the rotating wave approximation---varies only slightly on the same time scale of $\sim 1/\omega$, thereby reducing numerical demands for computing an approximation for the driven qubit's time evolution.  While the predicted time evolution agrees with the exact trajectory only once per drive period, mutually-disjoint sets of stroboscopically-defined points along the trajectory can be obtained by varying the gauge parameter $\beta_0=2\omega t_0$ with $\beta_0\in[0, 2\pi)$, which is a free parameter of our effective Hamiltonian $\H_\eff(t; t_0)$ through its dependence on $t_0$.  The ability to freely choose $\beta_0$ allows one to obtain the qubit state vector along the exact trajectory for any desired point in time.  The quantity $\beta_0$ is called a \emph{gauge parameter}, because when changing its value both the beginning and end points of the stroboscopic qubit trajectory are left invariant.  This statement, demonstrated empirically in Sec.~\ref{intro:gauge_kicks} (cf.~Fig.~\ref{gauge_and_kicks} and the animated time evolution$^{\ref{animation}}$), is exact---under the condition of convergence of the involved Magnus-Taylor expansions---since we have not made any approximation in our derivation.

While empirically we find that our series expression for the effective Hamiltonian $\H_\eff$ appears to converge vary rapidly for cases of practical interest, we cannot provide any formal guarantees about the convergence of this series; even the convergence of the Magnus expansion itself is a difficult subject \cite{blanes09}.  We have probed a speculation that, besides being defined by a series, our $\H_\eff$ may be given an axiomatic definition independent of the series analysis.  For the case of an analytic envelope $H_1(t)$, surely two of these axioms would be that 1)  $\H_\eff(t; t_0)$ is an analytic function of time $t$, and 2)  its time evolution operator agrees exactly with that of the rotating-frame Hamiltonian $\H_\rot$ at times $t_0$, $t_0+t_c$, $t_0+2t_c$, $\ldots$ These two axioms, however, are clearly not sufficient, because the exact Hamiltonian, $\H_\rot$, also satisfies them. Therefore at least one more axiom would be required.  

After examining various candidates for such an additional axiom, two of the authors have explored what seemed the most intuitive route.  When comparing the effective and exact Bloch-sphere trajectories for a driven qubit (see, e.g., Fig.~\ref{trajectories}), it is easy to note that the former trajectories traverse significantly shorter paths than the latter.  We argue that since the length of the qubit trajectory is related to the norm of the Hamiltonian via the time evolution operator, the positive eigenvalue of the effective Hamiltonian is likely smaller than that of the exact Hamiltonian.  Based on this, we considered the following third axiom, 3) for all analytic Hamiltonians $\H(t; t_0)$ satisfying axioms 1) and 2) stated above, the integral of the positive eigenvalue of that Hamiltonian, $\text{eig}_+(\H)$, taken over all times and the entire range of gauge parameters, $Q[\H] = \int_{-\infty}^{\infty} \d \tau \int_{0}^{2\pi} \d \beta_0 \, \text{eig}_+(\H)$, is minimized by $\H = \H_{\eff}$.  In Ref.~\cite{zeuch20} the validity of this third axiom has been tested by numerically minimizing the functional $Q[\H]$ for variable Hamiltonians.  This study, however, refutes our hypothesized third axiom by the identification of a counterexample, that is, a drive envelope has been found for which the functional $Q$ is minimized by a Hamiltonian $\H \neq \H_{\eff}$.  Nonetheless, we are still hopeful that a complete axiomatic definition may be discovered by identifying a suitable third axiom.  

If the drive envelope $H_1(t)$ is not a smooth function of time, the effective Hamiltonian needs to be supplemented by the kick operator formalism.  A kick operator is the generator of an impulse that connects effective trajectories before and after the times at which one of the envelope's derivatives, $\dersub{H_1}{k}(t)$, diverges in form of a $\delta$-function.  Expressing the kick operator as a series expansion in $1/\omega$ similar to that of the effective Hamiltonian, we have derived a systematic procedure to obtain these kick operators using the Magnus-Taylor expansion.  

Our method is based on resonant driving, for which the Hamiltonian contains a term \cite{evangelos_extension} consisting of an explicit periodic function (with frequency $\omega$) that is multiplied by an envelope function $H_1(t)$ [cf.~Eq.~(\ref{Hlab})].  For best convergence of our effective Hamiltonian series, the drive frequency $\omega$ is assumed large compared to the amplitude $H_1(t)$.  Indeed, as becomes clear from Fig.~\ref{trajectories}, if the amplitude of the drive envelope is similar to the drive frequency, or $|H_1(t)| \approx \omega$, many terms in the effective Hamiltonian series may be needed to achieve a high accuracy when approximating the time evolution.  We note that the presented theory is not designed to be applicable to arbitrary time-dependent Hamiltonians, although a similar approach for generic time-dependent perturbation theory may be conceivable.

The driven quantum two-level system is an excellent platform for introducing our effective Hamiltonian theory, because it allows for a visual presentation of the properties of our effective Hamiltonians and kick operators.  However, none of the steps taken within this work rely on the driven quantum system being two-dimensional, and under certain restrictions our work can be applied almost in parallel to Hamiltonians with more than one drive or with a time-dependent phase offset $\phi = \phi(t)$ in the Hamiltonian (\ref{Hrot0}).  We therefore envision that our basic theory can be extended to more generic problems including larger driven quantum systems or less restricted drives.  Furthermore, two-qubit gates of interest, for example the so-called cross-resonance gate, also involve resonant driving between two levels and thus could be effectively analyzed with our methods.  We are optimistic that our exact rotating wave approximation will have many applications in the forthcoming quest to more precisely analyze the creation of new, high precision logic gate operations for quantum computing.

\section*{Acknowledgments}

Many valuable insights for this work were gained through a collaboration on a related topic with Evangelos Varvelis.  Furthermore, useful discussions with Kai Segadlo, Veit Langrock, Alwin van Steensel and Cica Gustiani are gratefully acknowledged.  This work was supported by Intelligence Advanced Research Projects Activity (IARPA) under contract W911NF-16-0114.

\appendix

\section{Example Effective Hamiltonians}
\label{appendix:examples}

Assuming a completely analytic pulse envelope, here we give some concrete results.  As discussed in Sec.~\ref{procedure}, the effective Hamiltonian (\ref{Heff1}) of order $1/\omega^N$, as given in Eq.~(\ref{Hseries}), can be obtained by starting with the Hamiltonian (\ref{h0}) given in the rotating wave approximation, and then applying the recurrence relation (\ref{recursion}) repeatedly.  The examples shown below are for the generic rotating-frame Hamiltonian (\ref{Hrot0}) and various limiting cases of vanishing parameters $\phi$ and $\Delta$.  

We first write the effective Hamiltonians as a sum of terms $\H_i$, which are proportional to $1/\omega^i$ for $i=0$, 1, and 2,
\begin{equation}
		\H_{\eff}(t;t_0) = \mathcal H_0(t;t_0) + \mathcal H_1(t;t_0) + \mathcal H_2(t;t_0) + \mathcal{O}(1/\omega^3).
		\label{HSum}
\end{equation}
By convention, the dependence on $t_0$ is given through the dimensionless gauge parameter $\beta_0 = 2\omega t_0 \in [0,2\pi)$ introduced in Eq.~(\ref{beta0}).  

For the most generic case of arbitrary $\phi$ and $\Delta$ we find the lowest three terms of Eq.~(\ref{HSum}) to be
\begin{eqnarray}
		\H_0 &=& \frac{H_1}{4}[\cos\phi\sigma_x+\sin\phi\sigma_y]+ \frac{\Delta}{2}\sigma_z, \label{0} \\
		\H_1 &=& \frac{H_1^2}{32 \omega}(1-2 \cos(\beta_0+2 \phi))\sigma_z+ \frac{\Delta\ H_1}{8 \omega}[\cos(\beta_0+\phi)\sigma_x-\sin(\beta_0+\phi)\sigma_y] \nonumber \\
				&& + \frac{\dot H_1}{8 \omega}[\sin(\beta_0+\phi)\sigma_x+\cos(\beta_0+\phi)\sigma_y], 
				\label{leadingCorrection}\\
		\H_2 &=& \frac{H_1^3}{256 \omega ^2}[(-2 \cos\phi + 2 \cos(\beta_0+3 \phi)-\cos(2 \beta_0+3 \phi))\sigma_x \nonumber \\
				&& +(-2 \sin\phi + 2 \sin(\beta_0+3 \phi)+\sin(2 \beta_0+3 \phi))\sigma_y] \nonumber \\
				&& + \frac{\Delta H_1^2}{32 \omega ^2}(-1+\cos(\beta_0+2 \phi))\sigma_z - \frac{\Delta ^2 H_1}{16 \omega ^2}[\cos(\beta_0+\phi)\sigma_x-\sin(\beta_0+\phi)\sigma_y] \nonumber \\
				&& + \frac{3 H_1 \dot H_1}{32 \omega ^2} \sin(\beta_0+2 \phi) \sigma_z - \frac{\Delta \dot H_1}{8 \omega ^2}[\sin(\beta_0+\phi)\sigma_x+\cos(\beta_0+\phi)\sigma_y] \nonumber \\
				&& + \frac{\ddot H_1}{16 \omega ^2}[\cos(\beta_0+\phi)\sigma_x-\sin(\beta_0+\phi)\sigma_y].
		\label{2}
\end{eqnarray}
Note that here and below all temporal dependences of the Hamiltonians terms $\H_i = \H_i(t; t_0)$, the envelope $H_1 = H_1(t)$ and its derivatives $\dersub{H_1}{k} = \dersub{H_1}{k}(t)$ are all kept implicit.  

Setting $\Delta = 0$ in the above Hamiltonian terms, we find the effective Hamiltonian for on-resonant driving to be
\begin{eqnarray}
	\H_0 &=& \frac{H_1}{4}[\cos\phi\sigma_x+\sin\phi\sigma_y], \\
	\H_1 &=& \frac{H_1^2}{32 \omega}(1-2 \cos(\beta_0+2 \phi))\sigma_z+ \frac{\dot H_1}{8 \omega}[\sin(\beta_0+\phi)\sigma_x+\cos(\beta_0+\phi)\sigma_y], \\
	\H_2 &=& \frac{H_1^3}{256 \omega ^2}[(-2 \cos\phi + 2 \cos(\beta_0+3 \phi)-\cos(2 \beta_0+3 \phi))\sigma_x \nonumber\\
			&& +(-2 \sin\phi + 2 \sin(\beta_0+3 \phi)+\sin(2 \beta_0+3 \phi))\sigma_y]+ \frac{3 H_1 \dot H_1}{32 \omega ^2} \sin(\beta_0+2 \phi)\sigma_z \nonumber\\
			&& + \frac{\ddot H_1}{16 \omega ^2}[\cos(\beta_0+\phi)\sigma_x-\sin(\beta_0+\phi)\sigma_y].
\end{eqnarray}

Alternatively, setting $\phi = 0$ in the Hamiltonian terms (\ref{0})-(\ref{2}) yields
\begin{eqnarray}
	\mathcal H_0 &=& \frac{H_1}{4} \sigma _x+ \frac{\Delta}{2}  \sigma _z, \\
	\mathcal H_1 &=& \frac{H_1^2}{32 \omega}(1-2 \cos\beta_0)\sigma_z+ \frac{\Delta H_1}{8 \omega}[\cos\beta_0\sigma_x-\sin\beta_0\sigma_y] + \frac{\dot H_1}{8 \omega}[\sin\beta_0\sigma_x+\cos\beta_0\sigma_y],\\
	\H_2 &=& \frac{H_1^3}{256 \omega ^2}[(-2 + 2 \cos\beta_0-\cos(2 \beta_0))\sigma_x+(2 \sin\beta_0+\sin(2 \beta_0))\sigma_y]\nonumber \\
			&& + \frac{\Delta H_1^2}{32 \omega ^2}(-1+\cos\beta_0)\sigma_z - \frac{\Delta ^2 H_1}{16 \omega ^2}[\cos\beta_0\sigma_x - \sin\beta_0\sigma_y] + \frac{3 H_1 \dot H_1}{32 \omega ^2}\sin\beta_0\sigma_z \nonumber\\
			&& -\frac{\Delta \dot H_1}{8 \omega ^2}[\sin\beta_0\sigma_x+\cos\beta_0\sigma_y] + \frac{\ddot H_1}{16 \omega ^2}[\cos\beta_0\sigma_x-\sin\beta_0\sigma_y].
\end{eqnarray}

Finally, the Hamiltonian for the special case of both resonance and zero phase offset, $\Delta = 0$ and $\phi = 0$, is given by
\begin{eqnarray}
	\mathcal H_0 &=& \frac{H_1}{4}\sigma _x, \\
	\mathcal H_1 &=& \frac{H_1^2}{32 \omega}(1-2 \cos\beta_0)\sigma_z+ \frac{\dot H_1}{8 \omega}[\sin\beta_0\sigma_x+\cos\beta_0\sigma_y], \\
	\mathcal H_2 &=& \frac{H_1^3}{256 \omega ^2}[(-2 + 2 \cos\beta_0-\cos(2 \beta_0))\sigma_x+(2 \sin\beta_0+\sin(2 \beta_0))\sigma_y] \nonumber\\
			&& + \frac{3 H_1 \dot H_1}{32 \omega ^2} \sin\beta_0\sigma_z+ \frac{\ddot H_1}{16 \omega ^2}[\cos\beta_0\sigma_x-\sin\beta_0\sigma_y].
\end{eqnarray}

\bibliographystyle{elsarticle-num}  %\bibliographystyle{elsarticle-harv}
\bibliography{bibliography}

\begin{thebibliography}{10}
\bibitem{envelopeFunction}
The envelope shown in Fig.~\ref{trajectories}(c) is piecewise-defined as
  $H_1(t) = (H_1^{(\max)}/2) \left\{1 - \cos\left[ (1-a)H_1^{(\max)} t / a
  \right]\right\}$ if $0 \leq t/t_{\gate} \leq a$, $H_1(t) = H_1^{(\max)}$ if
  $a < t/t_{\gate} < 1-a$, and $H_1(t) = (H_1^{(\max)}/2) \left\{1 + \cos\left[
  (1-a)\left(H_1^{(\max)} t-\pi\right) / a\right]\right\}$ if $(1-a)t_{\gate}
  \leq t/t_{\gate} \leq a$ with $a= t_{\ramp} / t_{\gate} = 0.4$. The value for
  $t_\gate$ depends on the drive strength and is determined by the condition
  $\int_0^{t_\gate}\d t\, H_1(t) = 2\pi$ [cf.~Eq.~(\ref{NOT})].
\bibitem{H_discon}
The envelope shown in Fig.~\ref{fig:jumpop-trace}(a) is piecewise-defined as
  $H_1(t) = (H_1^{(\max)}/2)$ if $0 \leq t/t_{\gate} \leq 1/2$, and $H_1(t) =
  H_1^{(\max)}(3/2-t/t_\gate)$ if $1/2 < t/t_{\gate} < t_\gate$.
\bibitem{rotatedBS}
The coordinates $\theta$ and $\phi$ used in Fig.~\ref{fig:jumpop-trace}(b) are
  defined by $\theta = 2 \arccos\left(\left|\langle\psi|+\rangle\right|\right)
  - \pi/2$ and $\phi = \arg\left(\langle\psi|-\rangle\right) -
  \arg\left(\langle\psi| + \rangle\right)$.
\bibitem{evangelos_extension}
We note that an extension to the calculation of effective Hamiltonians of a
  qubit driven by multiple drives with equal periodicity has been considered in
  Sec.~3 in Ref.~\cite{varvelis19}.
\end{thebibliography}

\end{document}